\documentclass[preprint]{elsarticle}

\usepackage{amssymb,amsmath,graphicx,multirow,url}
\usepackage[utf8]{inputenc}

\journal{Information Sciences}

\urldef{\mails}\path|{lci, jmfluna, jhg}@decsai.ugr.es|
\newtheorem{thm}{Example}

\newproof{pf}{Proof}

\begin{document}

\begin{frontmatter}

\title{On the selection of the correct number of terms for profile construction: theoretical and empirical analysis\journal{Information Sciences}} %Selection of the correct number of terms to build parliamentary profiles for content-based recommendation

%The address of the authors Luis M. de Campos, J. M. Fernández-Luna and J.F. Huete is Universidad de Granada, Departamento de Ciencias de la Computaci\'on e Inteligencia Artificial, ETSI Inform\'atica y de Telecomunicaci\'on, CITIC-UGR, 18071 Granada, Spain and their email addresses                                 are \{lci,jmfluna,jhg\}@decsai.ugr.es }

\author[decsai]{Luis M. de Campos}
\ead{lci@decsai.ugr.es}

\author[decsai]{Juan M. Fern\'andez-Luna}
\ead{jmfluna@decsai.ugr.es}

\author[decsai]{Juan F. Huete\corref{cor}}
\ead{jhg@decsai.ugr.es}

\cortext[cor]{Corresponding author}

\address[decsai]{Departamento de Ciencias de la Computaci\'on e Inteligencia Artificial, \\ ETSI Inform\'atica y de Telecomunicaci\'on, CITIC-UGR, \\ Universidad de Granada, 18071, Granada, Spain}

\begin{abstract}
In this paper, we examine the problem of building a user profile from a set of documents. This profile will consist of a subset of the most representative terms in the documents that best represent user preferences or interests. Inspired by the discrete concentration theory we have conducted an axiomatic study of seven properties that a selection function should fulfill: the minimum and maximum uncertainty principle, invariant to adding zeros, invariant to scale transformations, principle of nominal increase, transfer principle and the richest get richer inequality. We also present a novel selection function based on the use of similarity metrics, and more specifically the cosine measure which is commonly used in information retrieval, and demonstrate that this verifies six of the properties in addition to a weaker variant of the transfer principle, thereby representing a good selection approach.

The theoretical study was complemented with an empirical study to compare the performance of different selection criteria (weight- and unweight-based) using real data in a parliamentary setting. In this study, we analyze the performance of the different functions focusing on the two main factors affecting the selection process: profile size (number of terms) and weight distribution. These profiles are then used in a document filtering task to show that our similarity-based approach performs well in terms not only of recommendation accuracy but also efficiency (we obtain smaller profiles and consequently faster recommendations).

\end{abstract}


\end{frontmatter}

%\maketitle
%\renewcommand{\shortauthors}{de Campos et al.}

\section{Introduction} \label{section:intro}

The original reason behind this paper was to meet two different information needs in a parliamentary context. Firstly, someone (e.g. a member of the public or a journalist) might need to find a certain politician to discuss a problem with them or simply interview them, respectively. In such cases, this person does not need any politician but rather one who works in the area connected with the matter concerned and who can attempt to solve the request or properly answer the interview questions. This is related to the expert-finding task \cite{BFS12}, which have been the subject of   substantial research over the past decade focusing on enterprise, online communities  or academic domains  \cite{Alha17,ribeiro17,tang08,WJA13}. Secondly, we can consider the situation whereby a new document (e.g. a press release, parliamentary initiative or a user's request or petition) is lodged by parliament. This document should then be filtered to those Members of Parliament (MPs) that might be interested in the information 
contained in the document, thereby solving the information overload problem. This concerns the document-filtering task or content-based recommendation task \cite{BOH13,RRS15}.
  
For both tasks, a structure such as a profile is needed to represent the topics of interest to the MPs. This profile might be created manually (where the users express explicitly their preferences \cite{balog08,BFS12} or using relevance feedback \cite{Alha17}) or automatically (by analyzing documents created during routine work \cite{BFS12,JIS16}). There are various reasons that advise against the  use manually-based methods \cite{BRB13}: people do not like having to complete long forms about their interests, they are generally biased towards current interests and are static since people rarely update their preferences. 

So, we would use automatic methods, as in our previous paper \cite{JIS16}, but considering  that a profile should be a {\em subset} of all the words used by the MPs in their speeches in different parliamentary  sessions.To tackle this problem, firstly we should apply any weighting scheme in order to determine for each used word its representativeness with respect to the MP preferences.  The output of this step is a list of words with their associated weight, possibly sorted.  Nevertheless, after this step  a selection process becomes necessary to separate the wheat from the chaff.  When a ranking of the terms is returned, it is possible to establish
a threshold in order to discard the   less relevant ones. Unfortunately, where to establish the threshold is not an easy-to-solve question \cite{BSA13}, and usually   heuristic approaches were used. So, the objective is to determine which are those terms (among all the terms in the vocabulary) that finally will belong to the MP's profile (those  truly representing her preferences). 
In this paper we will focus on this problem, motivated mainly by some findings from our previous research. 

Thus, in \cite{JIS16} we used the $N$-top ranked where $N$ is a constant parameter. We experimented with profiles of different fixed sizes and concluded that profiles with a large number of terms performed best. However, we realized that this selection scheme might not be entirely appropriate. One reason for this is that if an MP participates in more debates than another, then the number of possible keywords for selection is potentially greater. There are also MPs with more focused areas of political interest (they might, for example, specialize in areas such as education or employment) whereas others might have a much wider range of interests, and might therefore require a greater number of terms. The way MPs express themselves and their verbosity could also affect the length of their profiles. If small profiles are preferred, we could omit a large number of interesting terms to correctly represent those MPs with a wide range of interests. On the other hand, large profiles might include useless terms which 
would introduce noise into the recommendation time. The underlying hypothesis for this current research therefore states that the best results would be obtained if variable size profiles are used which are better adapted to each MP.

Finding a good selection scheme is not easy because two different factors need to be taken into account: firstly, the number of terms in the MP vocabulary, and secondly, the representativeness of each term, which depends on how the weights are distributed among the terms. What generally happens is that the more concentrated the weight distribution is, the fewer the terms required to represent the profile. Consequently, any useful scheme for evenly characterizing a profile should consider these two factors. Thus, the selection techniques considered in this paper can be classified into two groups: unweighted ones, which only consider the number of terms in the profiles, and weighted ones, which take into account the distribution of the weights among the terms. Each profile could have a potential different size depending on the representativeness of the selected terms.

In spite of these practical reasons for our interest, studying conceptual problems involved in the selection of the most representative items (in our case terms) from a ranked set of them is an important problem and one that also deserves to be studied from a theoretical point of view.  Correspondingly, it is possible to draw a parallel with the concentration theory \cite{egghe05} or the measurement of inequality in econometrics \cite{Atkinson70}. In this paper we will attempt to clarify the basic issues, to examine those properties that should be required for any employed function, and, on the basis of these, to discuss a new approach for tackling the selection problem based on a similarity measure. In the paper we analyze several approaches,  adapted to the problem at hand, with the aim of  identifying whether or not they satisfy the proposed principles. This analysis has provided us with insight into the expected behavior for each selection criteria in practice.
% 
% We have not overlooked our initial motive of selecting the most relevant terms from a document collection in order to build a profile. In this respect, it may be helpful to distinguish two objectives that arise from an empirical point of view: firstly, we are interested in determining those factors that potentially influence the selection process, and, secondly, we need to determine the quality of the different approaches under our original problem. In the empirical analysis we will use the same data considered in \cite{JIS16} from the Spanish Andalusian Parliament to create three  collections of documents of varying size and number.
% 
% In this context, our research will attempt to answer the following questions:
% 
% \begin{itemize}
% \item What principles should be required for a selection function?
% \item Can a similarity measure be a good alternative to guide a selection process?
% \item Given a document, which factors should affect the selection process?
%  \item Would it be better to use a variable approach whereby the selection criteria vary according to document size or weight distribution, or would it be preferable to use one which considers a set number of terms for every profile?
%  \item Is any one approach which is clearly better than the others?
%  \item What are the best parameters for the different methods?
%  \item Does the performance of these methods depend on the collection type?
% \end{itemize}

The rest of the paper is structured as follows. The next section describes the related work in the field of term selection. Section \ref{theoretical} presents those theoretical properties that should be satisfied by a selection function. Section \ref{sec:cutoff} analyzes five cutoff functions, including our novel approach, which are based on the use of similarity metrics. Section \ref{sec:exp1} presents an empirical study of these methods and shows how they perform when document-based profiles are used. We continue by focusing on a politician recommendation task in Section \ref{section:system} with a general introduction of the system used, details of the experimental design and the results obtained. We finish this paper with Section \ref{section:conclusions} with our conclusions about this research. The paper ends with an appendix which includes several demonstrations.

\section{Related work}

It is quite common in different disciplines to model a real world phenomenon by considering a set of attributes to describe the examples, cases, individuals or entities. The importance of each attribute is generally represented by a weight and it is quite common to use vector space models to represent this information. For instance, in a recent paper \cite{BSE17} on entity profiling, 60\% of the 52 papers reviewed, and which were published between 2000 and 2015, use such a representation. The advantages are that they are easy to implement, require much less learning work than other approaches and perform well on tasks that involve measuring the similarity between elements \cite{BRB13,TP10}. The range of disciplines can vary from econometrics (the attributes might be the workers and the weights their productivity), bibliometrics (where the items are the books/articles and the weights might be the number of times they are loaned out/cited) or, along the same lines as our research purposes, information retrieval (where the items are the words and the weights their occurrence in a given text), recommender systems (products and users ratings) or user modeling (weighting their preferences).

In terms of domains that use documents as input, vector space models have been used in areas such as search engines \cite{baeza11}, document classification or clustering \cite{forman03}, e-commerce \cite{RRS15}, research paper similarity \cite{HSC13}, enterprise information seeking \cite{Alha17,balog08}. In these domains, Zipf’s law \cite{zipf49} states that the frequency of a word is approximately proportional to the reciprocal of its rank in a frequency list. This law ensures dramatically skewed distributions for almost every language-applied statistic, and power scaling ensures that most words occur very infrequently. Nevertheless, this situation is often encountered in areas relating to information sciences: there are few prolific scientists and most publish only a few articles; there are few top reference journals for each scientific discipline but many journals occasionally publish related articles; a small group of movies are very popular and seen by many people whereas most are seen by only a few people. When building a vector space model in such domains, some common properties can therefore be identified: the size of the vector is quite large (it is related to the number of terms in the vocabulary, the number of users/items, the number of researchers, etc.); there is a severe sparse data problem since most attributes have relatively low weights (many with the value zero) and it is also quite normal for most of the informat

In theory, the fullest possible profile for an element would include every attribute in the domain and an unfeasibly large vector would be obtained. As \cite{Lowe01} says, in practice this is not a good approach since it is difficult for the end user to interpret its contents \cite{LZ06}. Identifying the most relevant attributes can help to improve system performance by reducing noise as well as saving computational resources \cite{forman03}. The simplest way to perform such a selection is to limit the number of vector components, keeping only the most representative ones \cite{TP10} which will help to improve system performance. For instance, in \cite{Lin98} the top-200 words are used to automatically construct thesauri showing that the number of comparisons needed greatly decreases while little precision is lost for this task. Similarly, in \cite{Lee08} different term weighting variants were used to select the most important N\% terms from each document. A threshold over a given metric is also used to select the distinctive bigrams from a list of all the possible combinations of two words occurring in the corpus \cite{MSC13}. In most cases, the final set of attributes has usually been obtained either heuristically or empirically but this task has hardly been studied.

Despite the approach used, this task concerns topics such as feature selection, feature extraction and dimensionality reduction, where the variables or features considered are the terms used in the transcripts of each MP's speeches.

Feature selection \cite{guyon03} is the process of selecting a subset of relevant features from the original set of features. However, feature extraction and dimensionality reduction techniques \cite{fodor02} (such as Principal Component Analysis and in some way topic based approaches as LDA \cite{Blei03}) create new features from functions of the original features by transforming the data in a high-dimensional space into a smaller dimensional space. In our case, and given that we shall use the selected terms in each profile to create documents to be used by an IRS, we are not interested in techniques that transform the features and so we focus on feature selection. Much of the work in feature selection is related with the field of classification. The three main categories of feature selection algorithms are wrappers, filters and embedded methods \cite{BSA13,CS14,guyon03,kohavi97}. Wrappers use the same model/classifier to score subsets of features according to their error rate. Filters score subsets of 
features using a measure that is independent of the chosen model. Embedded methods incorporate feature selection as part of the model construction process (and are specific to the given model). Since we do not rely on any given model in our case, we do not consider wrappers and embedded methods.

Within filter methods, we focus on the so-called feature ranking methods rather than explicitly searching for the best feature subset  \cite{BSA13,CS14}. Feature ranking methods rank individual features independently of the context of others, using some measure of the goodness, relevance or importance of each feature. In order to effectively perform feature selection, feature ranking methods need an additional criterion to determine a cutoff point in the ranking. The reason for discarding methods that search for the best feature subset (in addition to the greater computational cost) is that they tend to penalize redundant features. This is appropriate for classification problems. However, in our case avoiding redundant features (e.g. terms that frequently co-occur) is not necessarily a good idea since two redundant terms might enable matching with different documents.

When dealing with a classification problem, a number of well-known measures can be used to rank the features by computing scores between each candidate feature and the class variable. Examples of these measures include mutual information, information gain, $\chi^2$ statistic, odds ratio and Gini coefficient \cite{forman03,singh10,yang97}. Since we do not have the class variable, we are not faced with a (text) classification problem, and so these measures cannot be applied. Moreover the objective in our research is different, since we are looking for the most representative terms to be included in the profile (even if they are used frequently by several MPs), and not for those terms which are helpful for classification tasks, which seems to be the ones  distributed  most  differently  in  the sets of positive and negative examples of the class \cite{sebastiani02}. Instead we need to resort to such term importance measures used for indexing in the field of information retrieval as document frequency, term 
frequency or inverse document frequency \cite{anderson01,baeza11,moens06}. There is also related work about term selection in the context of query reformulation, for example to reduce long queries \cite{maxwell13} or to perform query expansion for relevance and pseudo-relevance feedback \cite{cao08}.

The question that we ultimately ask when building a selection function is whether a term subset is representative of the entire profile. In this paper we will see that the answer to this question is connected with the concentration of the term's weights in the profile. The concept of concentration of a distribution has been studied in many disciplines (e.g. econometrics, statistics, informetrics, ecology) under related concepts such as inequality, diversity, variation or dispersion of a distribution \cite{Atkinson70,egghe05,HH03,iszak96,PT82}. In most cases, the main purpose is to determine a measure which summarizes in a single value the entire distribution, such as for example the Gini index, coefficient of variation or Theil measure. It should be noted that these measures are mainly used to compare (or rank) several distributions. Although the concept of concentration is frequently used in the literature, in the paper we therefore introduce its use for designing a selection function. In this respect, and 
inspired by the work of L. Egghe \cite{egghe05,egghe10}, we will adapt those principles required for a concentration measure to our problem in the following section. These principles provide a foundation for an intuitive understanding of the selection problem, guiding the design and validation of any function used to determine the cutoff point.

\section{Theoretical foundations}
\label{theoretical}

Before discussing the properties, we will present some notation. Let $w(t,D_j)$ be a measure of the importance or weight of the term $t$ in a document $D_j$ within a document collection ${\bf D}=\{D_1,\ldots,D_N\}$. We then rank the terms in $D_j$ in decreasing order of importance to obtain an ordered list $L_j=(t_{1,j},t_{2,j},\ldots,t_{n_j,j})$, where $w(t_{i,j},D_j)\geq w(t_{i+1,j},D_j)$, $i=1,\ldots,n_j-1$ and $n_j$ is the number of different terms in $D_j$. We then select the first $l_j$ terms in this list, i.e. $\hat{D}_j=\{t_{1,j},\ldots,t_{l_j,j}\}$, where $l_j\leq n_j$ represents document $D_j$.

The question now is how we can determine the cutoff point $l_j$. If $l_j$ is too small, the resulting profile might not be representative enough of the user's real interests. Otherwise, if $l_j$ is too large, more space and time will be required to manage the profiles. Furthermore, it might also be also possible that the presence of many low weighted terms decrease the performance of a system. We therefore seek a cut function $C(L_j)$ that returns the cutoff point $l_j$ for a given ranking, thereby enabling us to select the most representative terms.

There are a number of similarities with the concentration theory \cite{egghe05}, which focus on the study of the degree of inequality or concentration in a set of positive numbers. In our case, these numbers are the term weights. It seems natural that such inequalities between weights should be taken into account when designing the selection function. If, for instance, there is a high number of terms with low weights and a low number of terms with high weights, then with only a small number of top-weighted terms can we can capture most of the distribution. One example of such a situation is the common 80/20 rule (or Pareto principle) which states that for many events 80\% of the effects (in our case, the weights) come from 20\% of the causes (the terms). Inversely, if all the weights are similar, a greater number of them are needed to capture the same proportion of the distribution. This is because in the first case the distribution is more concentrated than in the second.

This idea must be stated more rigorously. In general, therefore, if a distribution $L$ is more concentrated than a distribution $L'$, any function such that $C(L) \leq C(L')$, i.e. $l_j \leq l'_j$, can be considered a good alternative for determining the cutoff point. Nevertheless, many functions $C(L)$ can be found for our purpose and it is difficult to determine whether a given function is acceptable or not. For this purpose we formulate seven basic properties which have been borrowed from concentration theory \cite{egghe05} and which should be verified for any cutoff function. By analyzing these properties we shall obtain a deeper understanding of the behavior of any function that could be applied.

\begin{enumerate}
 \item[P1:] {\em $C(L_j)$ is minimal in the case of minimum uncertainty}, i.e. one term concentrates all the weight, $L_j=\{w_1,0,\ldots,0\}$, with $w_1>0$.

 \item[P2:] {\em $C(L_j)$ is maximal in the case of maximum uncertainty}, i.e. all the weights are equal, $L_j=\{w_1,w_2,\ldots,w_n\}$, with $w_i = w_j \forall i, j$ and $ w_i>0$

 \item[P3:] {\em invariant to adding zeros}. In other words, the addition of irrelevant terms to a document should not modify the cutoff point, i.e. $C(L_j)= C(L_j^*)$, where $L_j^*= L_j \cup \{0,\ldots,0\}$.
 
  In order to illustrate why this property is important, let us consider the case in which we have a single document and two alternative vector representations: in the first, we only consider the terms in the document and in the second, the vector dimension is equal to the size of the vocabulary, assuming as usual a zero weight for all the terms which do not belong to the document. In such a situation, if P3 does not hold, it is possible to obtain two different sets of selected attributes (profiles) associated with the same document, simply because we add irrelevant terms to the formal representation. Therefore, P3 is quite a reasonable property that should be expected for any cutoff function.

 \item[P4:] {\em invariant to scale transformation}, which indicates that the cutoff point should be independent of a scale factor (all the weights are multiplied by a common positive factor $k$), i.e. $C(L_j) = C(k\times L_j)$, where $k\times L_j=\{kw_1, kw_2, \ldots,kw_n\}$.

This is an important property since it guarantees that the cutoff point is not affected by a change in the unit of measurement. Thus, for instance, we can normalize the weights in a distribution (e.g. by dividing by the maximum or by the sum of all the weights) without affecting the cutoff point. For example, therefore, this axiom ensures that the cutoff point obtained using raw term frequencies as a weighting criteria has the same value as the one obtained if we instead consider their associated probabilities. This axiom also ensures that in situations where the weight is computed using logarithms, such as for example the cases of pointwise mutual information or inverted document frequency, changing the logarithm base does not affect the cutoff point. If this property is obeyed, it is possible for the cutoff point to be used to compare different distributions and it is not necessary to worry about the scale used to compute the weights or, otherwise, such a comparison would   be more difficult.

\item[P5:] {\em principle of nominal increase}, which indicates that if the weight of each term is increased by the same amount $h$, $h>0$, the distribution is less concentrated. Using econometric terminology, the wealth is better distributed (they represent a lower weight percentage) so the cutoff point should not be decreased, i.e. $C(L_j) \leq C( L_j+h)$, where $ L_j+h =\{w_1+h, w_2+h, \ldots,w_n+h\}$.

A common situation in which this principle has to be considered appears when estimating  multinomial distributions where the space event is unbounded, as is usually the case when working with text documents. Statistical approaches require the estimation of the probability for each word in a document  and, in the case of having limited data,  these estimations are not accurate enough. In such   situations, it is quite common to use Laplace smoothing for computing the  estimates, adding a constant positive value (pseudo-counts) to the raw frequencies before calculating the probabilities. The relative values of pseudo-counts represent the expected prior  probabilities. The larger its value, the greater the smoothing effect  (or in terms of concentration, the less concentrated is the resultant distribution). Therefore, P5 says  that smoothing the probabilities  should affect the cutoff point, moreover it ensures that more terms will be included in the profile (its size increases), as it could 
be expected. Note, that  using very high values of $h$  the resulting estimator will tend to the uniform  distribution, i.e. the situation with maximum uncertainty.

 \item[P6:] {\em transfer principle}, which states that given two weights $w_a,w_b$ with $w_a > w_b$, if weight is taken from the lowest weighted term, $w_b$, and given to the heavier one, $w_a$, then the weights are more concentrated and consequently the cutoff point should not be increased, i.e. $ C( L_j^+) \leq C(L_j) $, where $ L_j^+ =\{w_1,\ldots, w_a+h, \ldots,w_b-h, \ldots, w_n\}$.

 We shall now consider an example of a situation in which P6 applies. Let us suppose that we have a list of document terms and we can see that two represent the same high-level concept, such as for example two synonyms or two inflected forms of a word (different gender, number, tense, etc.). In this situation, it might be interesting to map both terms to the high-level concept (or map the inflected forms to their root, the objective of the stemming processes) and it is therefore necessary to compute the weight for this concept. In this case, and without losing generality, let us assume that term $w_a$, which appears most frequently in the document, is used to represent the concept (the root). We can therefore transfer the entire mass of weight (or a part of it) from the lowest weighted, $w_b$, to the canonical form. In this case, as P6 states, it would make sense that once this transfer has been made, fewer attributes are required to represent the same information (the cutoff point decreases) or, in other words, considering high-level concepts rather than terms ought to decrease the number of selected attributes.

It should be noted that any function that verifies P6 should also verify P1 and P2.

 \item[P7:] {\em richest get richer} inequality states that if the weight of the highest-weighted term is increased by an amount $h$, $h>0$, the distribution is more concentrated and therefore the cutoff point should not be increased i.e. $ C( L_j^*) \leq C(L_j)$, where $ L_j^* =\{w_1+h, w_2, \ldots,w_n\}$. It is worth mentioning that the same can be said if the {\em poorest source gets poorer} by discounting an amount, $h$, to the lowest weight.

One example of P7 in the poorest-gets-poorer version appears when the lowest term (or a set of terms) is removed from the original list possibly because this attribute is considered irrelevant for research purposes. In terms of weights, this is equivalent to 'artificially' setting its weight to zero, i.e. $h=w_n$. In such situations, since the weight's distribution changes, an impact on the cutoff point might be expected. The distribution becomes more concentrated and therefore if P7 holds, we can guarantee that the cutoff point might decrease; otherwise, behaviour will be considered counterintuitive.

 \end{enumerate}

\section{Cutoff functions} \label{sec:cutoff}

In this paper we will consider two different approaches for selecting the cutoff function: the weight-oriented approach and the unweight-oriented approach. We will analyze both by considering whether they obey the above properties in such a way that the more properties a selection criteria has, the better the quality of the terms selected.

%
% For illustrative purposes, Figure \ref{figure:terminos} shows the weights (y-axes) for the first 162 terms (the most representatives ones) for four different documents\footnote{These documents represent different user profiles learned under different conditions (see Section \ref{section:experiments}).}. It should be noted that although we only show the first 162 terms of each one, there are great differences between the total number of terms and so, for example the document {\em I-Sup(a)} has only 162 terms whereas {\em M-Prof} contains 4560 terms.
%
% \begin{figure} [ht]
% \begin{center}
% \includegraphics[width=0.90\textwidth]{Imagenes/perfilesFrecuencia.jpg}
% \end{center}
% \caption{Weights of the terms in four different documents (profiles), ranked in decreasing order of weight} \label{figure:terminos}
% \end{figure}

\paragraph{Unweight-oriented:} In this case, the terms are selected without considering how representative they are for the document. Although these criteria are to some extent blind, they are easy to understand on an intuitive level.
\begin{description}

\item[FN:] The simplest way to select $l_j$ is to consider a {\it fixed number} of terms, $m$, i.e. to select the $m$ most important terms, -- in this case, $l_j=\min(m,n_j)$. Examples of the many papers that use this approach are \cite{allan99,buckley94,nanas04,verbene16}. This is a `one size fits all' solution and its effectiveness might be problematic. Nevertheless, there are situations where this approach might be useful, such as, for example, if we want to compare the performance under similar circumstances of different weighting criteria, i.e. different term rankings.

Whether or not this approach verifies some of the required properties is circumstantial in that it neither depends on the number of terms nor the distribution of the weights. We will not therefore discuss the properties for this approach.

\item [FP:] Another simple way that enables a variable number of terms for each document is to consider a {\it fixed percentage}, $per$, of terms. Although this function might be considered to be a good indicator of the amount of information contained in the profile, it only considers one of the factors that influences the concentration: the number of terms. The cutoff point then becomes $l_j=\mbox{round}(n_j*per/100)$. Usage examples for this method are \cite{toucedo10}, in the context of patent retrieval, and \cite{tsegay09}, in the context of snippet generation.

We will now analyze which properties are satisfied. Properties P1 and P2 are not verified trivially. Moreover, FP does not verify P3 since the addition of irrelevant terms, which provide no relevant information about the content of document $D_j$ , increases the value of $l_j$. This is considered to be an undesirable property in related tasks, such as for example the definition of similarity \cite{ahlgren03} or concentration \cite{egghe05} measures. Properties P4 to P7 are verified trivially since the criteria do not consider the weight values and so the cutoff point is unaffected by changes in distribution. We consider this to be a circumstantial fulfillment.
\end{description}

\paragraph{ Weight-oriented:} In this case, the term weights are used as a guide to determine the final cutoff point. We will present two approaches that use the weights but they do not take into account the number of terms in the original ranking. Once again, therefore, only one factor is considered.

\begin{description}
\item[VT:] The first alternative should be to select those terms with a higher weight than a fixed threshold $\delta$ (see for example \cite{evans95} in the context of query expansion), -- in this case $l_j=\max(i \,|\, w(t_{i,j},D_j)\geq \delta)$. Another example of this technique is \cite{rojas07}, where the specific measure $w$ being used is entropy (which does not depend on the document being considered) and the threshold $\delta$ is a fixed percentage of the maximum entropy of all the terms.
We believe that fixing a common threshold $\delta$ for all the documents is quite difficult and possibly problematic as it does not take into account any transformation on the scale (P4) as the following example shows. Given $L1=\{10,7,5,3,2,1\}$ and $L2 = \{ 1.0, 0.7, 0.5,0.3,0.2,0.1 \}$ and a value $\delta=4$, then $l_1 = 3$ and $l_2=0$, i.e. the first three terms were selected for $L1$ and none for $L2$.

Following \cite{rojas07} we consider using a {\it variable threshold} $\delta_j$ which depends on the document $D_j$, using a percentage $per$ of the maximum weight of the terms in $D_j$, $\max_j=w(t_{1,j},D_j)$. We then define the variable cutoff point as $l_j=\max(i \,|\, w(t_{i,j},D_j)\geq \max_j*per/100)$. This is equivalent to normalizing all the weights by dividing by $\max_j$ (in the example below $L2$ was obtained after normalizing $L1$) and then selecting those terms with a normalized weight which is greater than a given threshold, $\delta$, with $\delta \in [0,1]$. Following on with our example, if we set $\delta=0.4$ ($per=40\%$) we obtain $l_1=3$ and $l_2=3$, which is more natural.
Nevertheless, one of the drawbacks of this approach is that it heavily depends on the weight of the most representative term in the document.
%Thus, considering our example, in the case of {\em I-Sup(a)}, using $per = 50\%$ the final profile will only contain four terms, $l=4$.

We will now examine whether this strategy satisfies the different properties. Properties P1 and P2 hold trivially. Properties P3 and P4 also hold because the cutoff point is unaffected once the the distribution has been transformed. If we focus on the principle of nominal increase, while shift transformations affect the cutoff point P5 is still obeyed (proof is trivial). For example, if we construct $L3$ by adding one to all the weights in $L2$, when $\delta = 0.4$, $l_3=6$, i.e. all the terms in $L3$ were selected. Property P6, however, does not hold as the following example shows: assume that $L4 = \{ 1.0, 0.7, 0.5, 0.4, 0.1, 0.1 \}$, i.e. an amount of $0.1$ was transfered from $w_5$ to $w_4$ in $L2$. In this case, for $\delta = 0.4$, $l_4 = 4 > 3 = l_2$, increasing the cutoff point. Finally, P7 holds trivially.

\item[RC:] Another approach, in an attempt to avoid changes in weights after a shift transformation, i.e. a nominal increase, is the so-called {\it range-based term cutoff} \cite{stoica00}, used in the context of term selection in learning a query from examples. Let $\min_j=w(t_{n_j,j},D_j)$ be the minimum weight of the terms in $D_j$, and let $per$ be a given percentage. The cutoff point in this case is therefore $l_j=\max(i \,|\, w(t_{i,j},D_j)>\min_j+per/100*(\max_j-\min_j))$. The idea behind this method is to attempt to determine the rank of the term after which the curve of term weights levels out, under the assumption that terms in the flat part of the curve are not useful (for increasing precision). It is worth mentioning that in the case of $\min_j =0$, {\it range-based term cutoff} and {\it variable threshold} are the same approach.

Focusing on the fulfillment of the different properties we can see that P1 and P2 are verified. For P3, since the addition of irrelevant terms could modify $\min_j$ and therefore the cutoff point, this property does not hold. This approach is invariant to scale and shift transformations and consequently P4 and P5 hold (P5 more strictly so since the cutoff point remains unchanged). With the same arguments as those used in VT, we can see that P6 does not hold and finally P7 holds trivially.
\end{description}

\subsection{A new approach for term selection} \label{section:SC}

In this section we will propose another method for selecting terms based on the idea of similarity between documents. Given a document $D_j$ and its ranked list of terms $L_j=(t_{1,j},t_{2,j},\ldots,t_{n_j,j})$, let us consider the sub-documents $D_j^i$ of $D_j$ comprising the first $i$ terms in this list (and their corresponding weights), i.e. $D_j^i=\{t_{1,j},\ldots, t_{i,j}\}$. Let $Sim$ be a similarity measure between documents and let us consider the similarities $Sim(D_j^i,D_j)$, $i=1,\ldots,n_j$. Obviously $Sim(D_j^i,D_j)\leq Sim(D_j^k,D_j)$ if $i<k$ and $Sim(D_j^{n_j},D_j)=1$ (as $D_j^{n_j}=D_j$). Given a fixed percentage $per$, our proposal is therefore to determine the cutoff point $l_j$ as $l_j=\min(i \,|\, Sim(D_j^i,D_j)\geq per/100)$. This means that we select (in decreasing order of weight) as many terms as necessary to obtain a similarity with the original document greater than a given percentage.

We will call this the {\it similarity-based term cutoff} approach (SC) and it depends on the definition of the similarity measure $Sim$. For example if we use the cosine similarity, then

$$
Sim(D_j^i,D_j)= \frac{\sum_{t\in D_j^i\cap D_j}w(t,D_j^i)w(t,D_j)}{\sqrt{\sum_{t\in D_j^i}w(t,D_j^i)^2}\sqrt{\sum_{t\in D_j}w(t,D_j)^2}}.
$$

Assuming that $w(t,D_j^i)=w(t,D_j)$ if $t\in D_j^i\cap D_j$, then we obtain

$$
Sim(D_j^i,D_j)= \frac{\sum_{k=1}^i w(t_{k,j},D_j)^2}{\sqrt{\sum_{k=1}^i w(t_{k,j},D_j)^2}\sqrt{\sum_{k=1}^{n_j} w(t_{k,j},D_j)^2}},
$$

\noindent which simplifies into

$$
Sim(D_j^i,D_j)= \sqrt{\frac{\sum_{k=1}^i w(t_{k,j},D_j)^2}{\sum_{k=1}^{n_j} w(t_{k,j},D_j)^2}}.
$$

When the cosine measure is used, our approach is similar to the method used in \cite{lewis94} in the context of text classification, which computes the sum of the weights of all the terms and selects terms in decreasing order of their weights until a specified fraction of the sum has been achieved.
%
% \begin{figure} [t]
% \begin{center}
% \includegraphics[width=0.80\textwidth]{Imagenes/transfer.eps}
% \end{center}
% \caption{Graph representing the plot of different curves obtained using $Sim(D^i,D)$: the higher the curve, the more concentrated the distribution} \label{figure:transfer}
% \end{figure}

\begin{figure}[t]
 \centering{\includegraphics[width=0.7\textwidth]{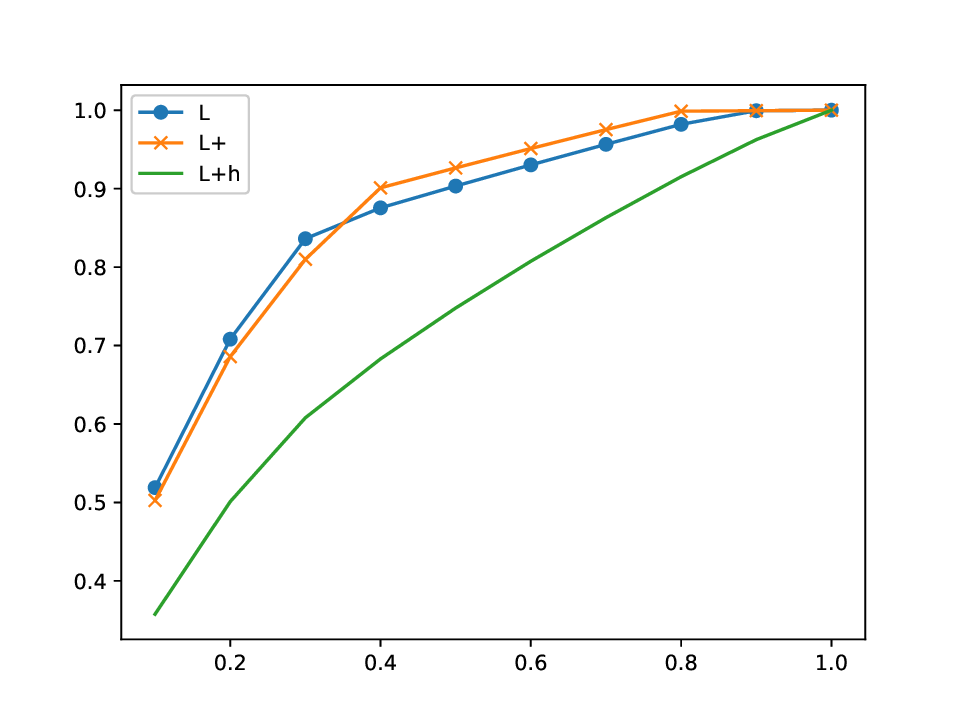} }

\caption{The figure shows a concentration curve for the distributions $L= \{14, 13, 12, 7, 6, 5, 5, 5, 5, 1\}$, $L+=\{14, 13, 12, 11, 6, 5, 5, 5, 1, 1\}$ and $L+h=\{54, 53, 52, 47, 46, 45, 45, 45, 45, 41\}.$}
\label{figure:analysis}

\end{figure}

A similar graph to the Lorenz curve can be used to illustrate how the proposed similarity varies in relation to the parameter $i$ (see Figure \ref{figure:analysis}). The percentage of selected terms is plotted on the x-axis, i.e. $i/n$, and the cosine similarity, i.e. $Sim(D^i,D)$, on the y-axis. Since the weights are considered in decreasing order, each distribution is represented on the graph as a concave polygonal curve that increases until the point (1,1). This graph displays the concentration of each distribution: the higher the curve, the more concentrated the distribution is.

%
% \begin{figure}[t]
% \caption{The figure shows the concentration curves and the table summarizes the cutoff functions satisfying the proposed principles. A bullet ($\bullet$) means circumstantially verified and a check mark ($\checkmark$) represents verified.\label{figure:analysis}}
% \centering{
% \begin{tabular}{cc}
% \begin{minipage}{0.65\textwidth}
% \includegraphics[width=0.95\textwidth]{Imagenes/transfer.eps}
% \end{minipage}
% &
% \begin{tabular}[b]{|c|c|c|c|c|c|c|c|}
% \multicolumn{8}{c}{Principles satisfied}\\
% \hline
% & P1 &P2 & P3 & P4 & P5 & P6 & P7\\
% \hline
% FP & & & $\bullet$ & $\bullet$ & $\bullet$ & $\bullet$ & \\
% VT & $\checkmark$ & $\checkmark$ & $\checkmark$ & $\checkmark$ & $\checkmark$ & & $\checkmark$ \\
% RC & $\checkmark$ & $\checkmark$ & & $\checkmark$ & $\checkmark$ & & $\checkmark$ \\
% SC & $\checkmark$ & $\checkmark$ & $\checkmark$ & $\checkmark$ & $\checkmark$ & wP6 & $\checkmark$ \\
% \hline
%
% \end{tabular}
% \end{tabular}
% }
% \end{figure}

We will now examine whether the different properties are verified by the similarity-based approach. Property P1 holds trivially and P2 also holds because the curve raised the minimum possible concentration. Property P3 holds trivially, since the addition of non-relevant terms does not modify the cosine measure. Generally speaking, cosine is invariant to scaling (so P4 holds) but not to shift transformations. Nevertheless, the principle of nominal increase (P5) is a weaker property because it does not require strict equality. Property P5 will be satisfied if the distribution $L+h$ is less concentrated than $L$ (as is the case of the distribution $L+h$ in Figure \ref{figure:analysis} which has been obtained after nominally increasing all the weights in $L$ by 40). In order to check that P5 holds, $Sim(D^i,D) \geq Sim((D+h)^i,(D+h))$, must be verified for each $i$. We should mention that $P5$ does not generally hold for the cosine measure, i.e. given any two arbitrary weighted vectors (see \cite{egghe10}). 
Nevertheless, in our case, $D^i$ is a subvector of $D$, thereby enabling us to guarantee that the nominal increase principle is verified (see   \ref{sec:demoP5}).

Finally, if we focus on the transfer principle, P6 does not hold for the cosine measure, as the following example shows:

\begin{thm}
  Consider the distribution $L$ in Figure \ref{figure:analysis} and assume a threshold $per=82\%$. In this case, the cutoff point is $l_j= 3$ since $Sim(D^2,D)=0.7081$ and $Sim(D^3,D)=0.8362$. Let $w_a =7$ and $w_b=5$ and let $L+$ be the distribution obtained after an amount of 4 was transferred from $w_b$ to $w_a$, $L+$ in Figure \ref{figure:analysis}. If we consider the same threshold, $l^+_j = 4$ since $Sim(D^{+3},D^+)= 0.8099$ and $Sim(D^{+4},D^+)=0.9010$.
\end{thm}

If we examine the $L$ and $L+$ curves in Figure~\ref{figure:analysis}, we will see that they intersect once, verifying that before the index associated to the value $w_a$ that receives the transfer, the distribution $L+$ is less concentrated than the distribution $L$, whereas after this point distribution is more concentrated. This is because the norm of the document associated to $L^+$ is greater than the norm of the document associated to $L$, i.e. $||D^+|| > ||D||$, penalizing the similarity value for those terms in positions $l < l_a$. In any case, we believe that this behavior is significant since it reflects the fact that weight transfer affects overall distribution and this idea is captured by cosine similarity. The following weaker transfer principle can therefore be formulated:

\begin{itemize}
\item[wP6] {\em weak-transfer principle} which states that given $ L_j =\{w_1,\ldots, w_n\}$ and two weights $w_a,w_b$ with $w_a > w_b$ and given $ L_j^+ =\{w_1,\ldots, w_a+h, \ldots,w_b-h, \ldots, w_n\}$ obtained by transferring an amount of weight from the lowest term, $w_b$, to the heaviest one, $w_a$. Let $l^+_a$ be the position of $w_a+h$ in $L_j^+$, then the weights are less concentrated for $l<l^+_a$ and more concentrated for $l\geq l^+_a$. The cutoff point should consequently increase before $l^+_a$ and decrease after $l^+_a$, i.e.

\begin{eqnarray}
\mbox{if } C(L_j)< l^+_a \mbox{ then } C(L_j) \leq C( L_j^+) \nonumber \\
\mbox{if } C(L_j)\geq l^+_a \mbox{ then } C(L_j) \geq C( L_j^+) \nonumber
\end{eqnarray}
\end{itemize}

This property indicates that the effect of a transfer to the cutoff point shall depend on the final position of the receiving term ($w_a$) in the new ranking. Using the example used in P6, if, following the transfer, the semantic concept receives enough weight to be placed among the most important ones (those below $C(L_j)$), the number of selected attributes might therefore be reduced, otherwise this need not be true.
 
In   \ref{sec:demoP6} we will demonstrate that cosine similarity verifies wP6. Finally, in terms of the richest gets richer principle, the inequality in P7 is also verified by cosine measure (proof is also included in   \ref{sec:demoP7}).

To conclude this section, Table \ref{tab:analysis} summarizes the different cutoff functions in terms of the verified properties. From this table, we can conclude that cosine measure is a good concentration measure and can therefore be used to determine the cutoff point in our framework.

\begin{table}[t]
\caption{Summary of the principles satisfied by the cutoff functions. A bullet ($\bullet$) means circumstantially verified and a check mark ($\checkmark$) represents verified.\label{tab:analysis}}
\begin{center}
\begin{tabular}[b]{cccccccc}
\multicolumn{8}{c}{Principles satisfied}\\

 & P1 &P2 & P3 & P4 & P5 & P6 & P7\\
 \hline
 FP & & & & $\bullet$ & $\bullet$ & $\bullet$ & $\bullet$\\
 VT & $\checkmark$ & $\checkmark$ & $\checkmark$ & $\checkmark$ & $\checkmark$ & & $\checkmark$ \\
 RC & $\checkmark$ & $\checkmark$ & & $\checkmark$ & $\checkmark$ & & $\checkmark$ \\
 SC & $\checkmark$ & $\checkmark$ & $\checkmark$ & $\checkmark$ & $\checkmark$ & wP6 & $\checkmark$ \\
 
\end{tabular}
\end{center}
\end{table}
%
% \paragraph{NO includio en el trabajo}
% Otras medidas que satisfacen principio de nominal increase son Jaccard Index y Dice \cite{egghe10}
% \[J(X,Y) = \frac{\sum xy}{\sum x^2 + \sum y^2 - \sum xy}\]
% \[ Dice = \frac{ 2\sum xy}{\sum x^2 + \sum y^2 }\]

\section{Cutoff points for document-based collections} \label{sec:exp1}

Once we have studied the theoretical properties of the cutoff functions, we will study their performance in practice by considering a real problem: determining the cutoff point for building profiles in a parliamentary framework. As we said, the difficulty in defining a cutoff function arises because it comprises two separate components: the \texttt{size} of the ranked list of terms, $L$, and the \texttt{distribution} of the weights in $L$. The cutoff function attempts to combine both components into a single value in such a way that the selected terms properly characterize the entire list. In this section, we will experimentally explore the relationships between these two components when considering real data.

\subsection{Source collections}
% In previous work \cite{JIS16} we built various types of MP profiles. In each case, these were compiled from the transcriptions of their parliamentary speeches when discussing initiatives\footnote{An initiative consists of the parliamentary discussion about a request presented by an MP or a political group. In addition to general information such as the title and date, an initiative also contains the sequence of speeches of each MP participating in the discussion. In particular, we consider the Records of Parliamentary Proceedings from the Spanish Andalusian Parliament for the 8th term of office (http://irutai2.ugr.es/ColeccionPA/legislatura8.tgz).} collected in the records of the parliamentary proceedings.

In order to test the cutoff functions under several circumstances, we have designed three different types of source collections (s-collections) from which the profile documents are built (as in \cite{JIS16}), varying the number of documents and their sizes:

\begin{itemize}
 \item Collection of every intervention (A-Col): a virtual document is created for each MP containing all of her speeches  when discussing initiatives\footnote{ In particular, we consider the Records of Parliamentary Proceedings from the Spanish Andalusian Parliament for the 8th term of office (http://irutai2.ugr.es/ColeccionPA/legislatura8.tgz).},  so the number of documents is the same as the number of MPs. Related to the document length, it  is the largest of the three collections presented.

 \item Collection of committee interventions (C-Col): since MPs participate on different committees relating to specific areas of interest (agriculture, education, health, economy, etc.) various virtual documents are created for each MP containing their speeches on the different commissions. If an MP is involved on different committees, one document for each committee will be created containing their speeches. This new collection increases the number but decreases the size of the documents.

\item Collection of single interventions (I-Col): the documents in this collection are created on the basis of initiative. There is one document for  each initiative in which the MP is involved. If, for example, the MP intervenes in ten initiatives, then ten documents will be created containing the corresponding speeches. The total number of documents is therefore larger than committee-based ones but they are smaller in size.
\end{itemize}

  The documents contained in these s-collections will be referred to as s-documents. In Table \ref{table:ColSizes1} we present the number of documents in each s-collection and their average sizes in terms of the number of keywords in the weighted ranking $L$, as well as the standard deviations.

\begin{table}[tbp]
	\centering
	\caption{Sizes of source collections\label{table:ColSizes1}}
	\begin{tabular}{cccc} \hline
			& \textbf{A-Col} & \textbf{C-Col} & \textbf{I-Col} \\
			 \textbf{ Number of docs} & 316 & 1192 & 10023 \\
			 \textbf{ Average size} & $1900.30 \pm 2108.77$ & $1028.57 \pm 1103.72$ & $321.27 \pm 216.75$ \\ \hline
	\end{tabular}
	
\end{table}

\subsection{Weighting measures} \label{sec:w}

In this paper we will explore four different alternatives for computing the weights of the terms:

\begin{itemize}
\item {\it Term Frequency} (TF)which captures the idea that the most representative terms for a document are the ones most frequently used, i.e. we use the raw frequency of term $t$ in document $D_j$, denoted by $f_{tj}$.
$$
TF(t,D_j)=f_{tj}.
$$
\item {\it Term Frequency-Inverted Document Frequency} (TFIDF) \cite{baeza11}, where relevant terms are those that occur frequently in the corresponding document (i.e. TF is high) but the term rarely occurs in other documents in the corpus   (i.e. IDF is high). IDF is measured as the logarithm of the ratio between the number of documents in the collection, $N$, and the number of documents containing the term $t$, i.e. $N_t$.

$$ 
TFIDF(t,D_j) = f_{tj}\times \log(N/N_t).
$$

\item {\it Pointwise Mutual Information} (PMI) \cite{BL07,TP10} is  designed to give high weights when there is an relevant relation between the term and the document.  PMI therefore compares the probability that a term occurring in document $D_j$ in terms of the one expected if we consider that both the document and the term are statistically independent. If $M$ is the total number of term occurrences in the collection, it therefore follows that

$$
PMI(t,D_j) = \log{\frac{f_{tj}/M}{(\sum_j f_{tj})/M\times (\sum_t f_{tj})/M}}.\\
$$

If there is a relationship between $t$ and $D_j$, we should expect $f_{tj}/M$ to be larger than it would be if it were independent, $(\sum_j f_{tj})/M \times (\sum_t f_{tj})/M$, and consequently PMI is greater than zero. Otherwise, if the term is unrelated, PMI is less than zero.

%PPMI is a variation in which all terms with PMI less than zero  are replaced with zero.
% 
% $$
% PPMI(t, D_j) = \left\{ \begin{array}{ll}
%                        PMI(t,D_j) & if   PMI(t,D_j) > 0\\
%                        0 & \mbox{otherwise}
%                       \end{array}
% \right.
% $$

 \item {\it Difference} (Diff)  was introduced in \cite{KBS16} in the context of personalized search and was one of the weighting schemes used in \cite{JIS16} for building MP profiles, and has certain similarities to the relative document frequency proposed in \cite{nanas04}. The Diff measure of a term $t$ for a document $D_j$ is the normalized frequency of $t$ in $D_j$ minus the normalized frequency of $t$ outside $D_j$ (i.e. in the other documents in the collection) and is computed as:

$$
Diff (t,D_j)=\frac{f_{tj}}{\sum_t f_{tj}} - \frac{\sum_{k \neq j} f_{tk}}{M-\sum_t f_{tj}}.
$$

\end{itemize}

One advantage of Diff and PMI measures is that they have a natural cutoff: if $w_t \leq 0$, it is because the term is more likely to occur in the context (collection) than in the document ($D_j$), therefore $t$ is not a representative term of $D_j$. More specifically, if the weight of any term is less than zero, it is replaced with zero. Throughout this paper we will use this idea for both Diff and PMI, which in the case of PMI is called positive pointwise mutual information (PPMI). This allows us to remove approximately 20\% to 30\% of the terms in a document depending on the collection used. Nevertheless, the number of selected terms is large and, as we will see, a further cutoff becomes useful.

\subsection{Analyzing weighted-oriented cutoff functions}\label{sec:cv}

\begin{figure}[tb]
%I histogramaCV
\centering{
\includegraphics[width=0.6\textwidth]{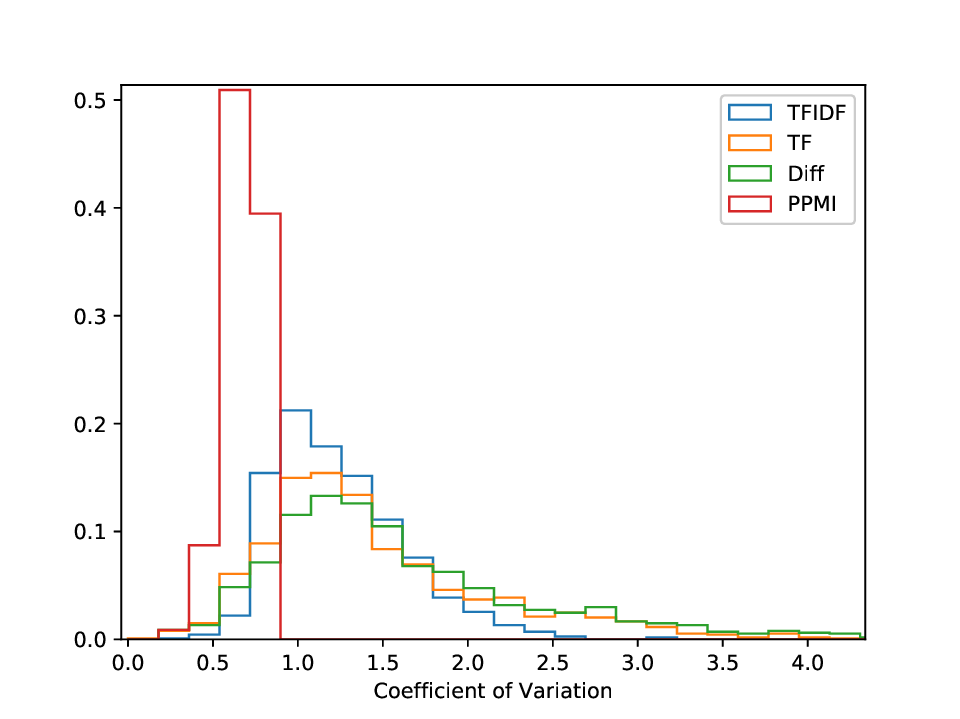} 
\caption{Coefficient of Variation for the different weighting criteria using C-Col.}
\label{figure:CV}

}
\end{figure}

As we mentioned previously, our approach for determining the cutoff point is related to the concentration of the term's weights in the document. We expect that the more unequal the distribution, the fewer number of terms needed to represent the profile. The concentration will depend on the weighting measure used, and there are various differences between them, as Figure \ref{figure:CV} shows (for the sake of clarity we only present the results for C-Col, although the trends are similar for A-Col and I-Col). In this histogram, the x-axis represents the coefficient of variation (CV), a well-known concentration measure of a distribution. It should be remembered that the CV measures the dispersion of the weights in terms of the mean and that the larger the CV, the more variable (diverse or concentrated) the data. On the other hand, the y-axis shows how probable the value is. Since CV is independent of the unit in which the measurement has been taken, we are able to compare the weighting functions used. We can therefore obtain an ordering among the different measures: PPMI $ \prec $ TFIDF $\prec$ TF $ \prec$ Diff, where $a \prec b$ means that the measure $a$ obtains more equal distributions than measure $b$. It is interesting to note that for TF, TFIDF and Diff less than 20\% of the distribution has a CV of below 1.0 whereas this is true for all the distributions obtained using PPMI.

\begin{figure}[t]

\centering{
\includegraphics[width=0.49\textwidth]{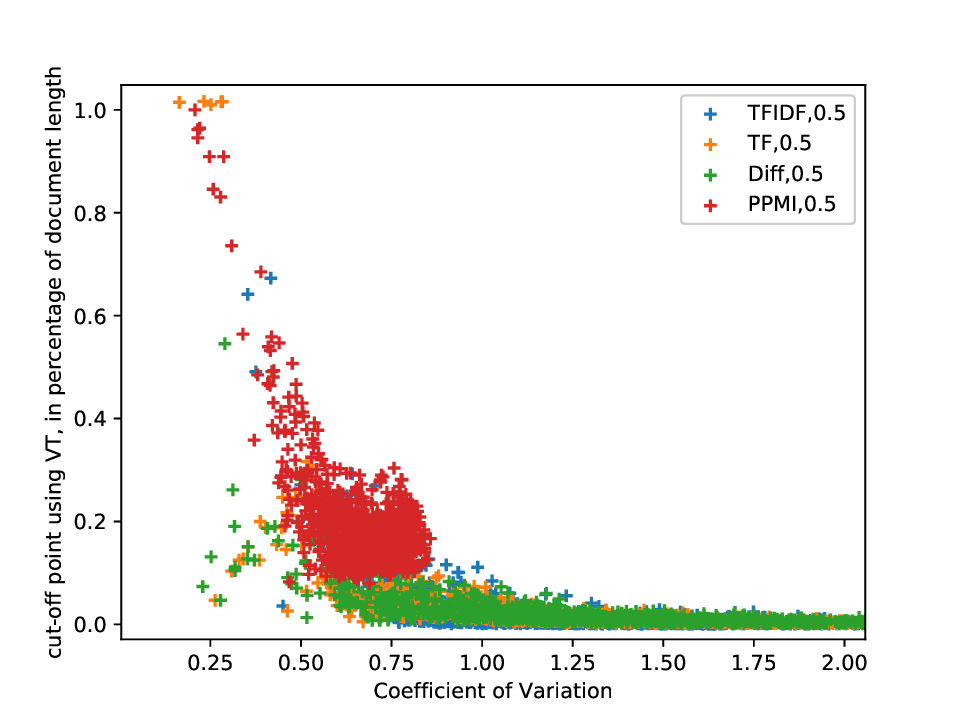} \includegraphics[width=0.49\textwidth]{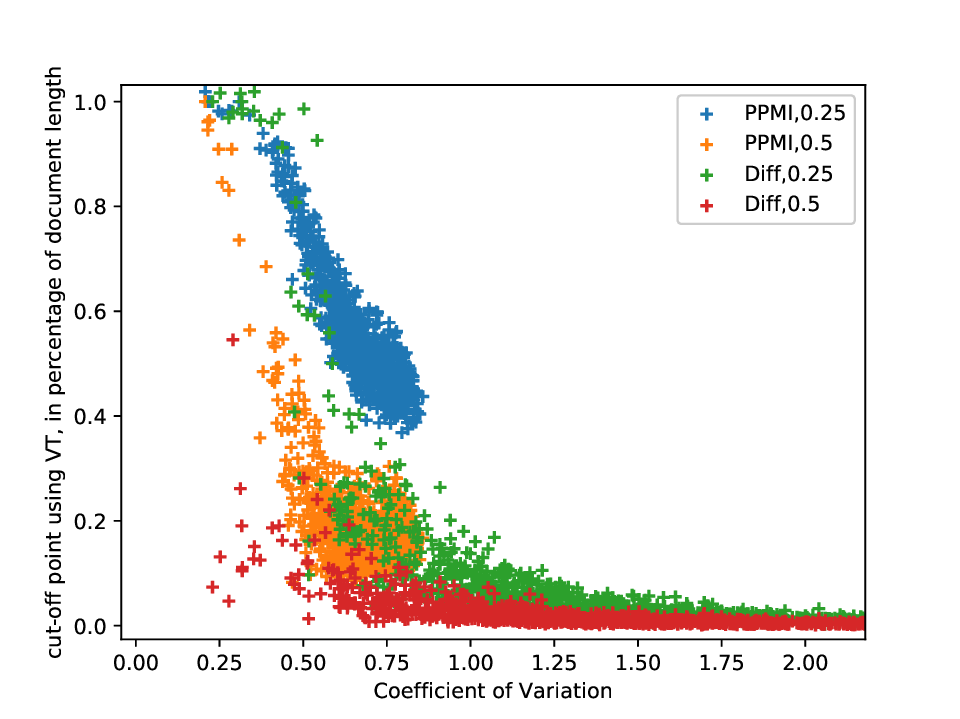}
\includegraphics[width=0.49\textwidth]{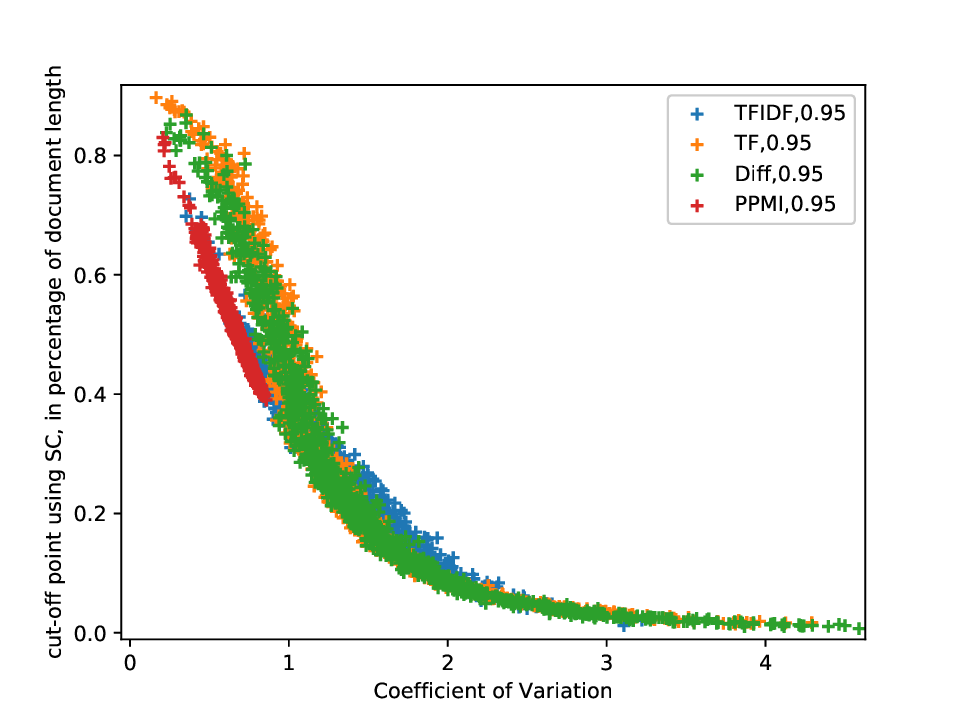} \includegraphics[width=0.49\textwidth]{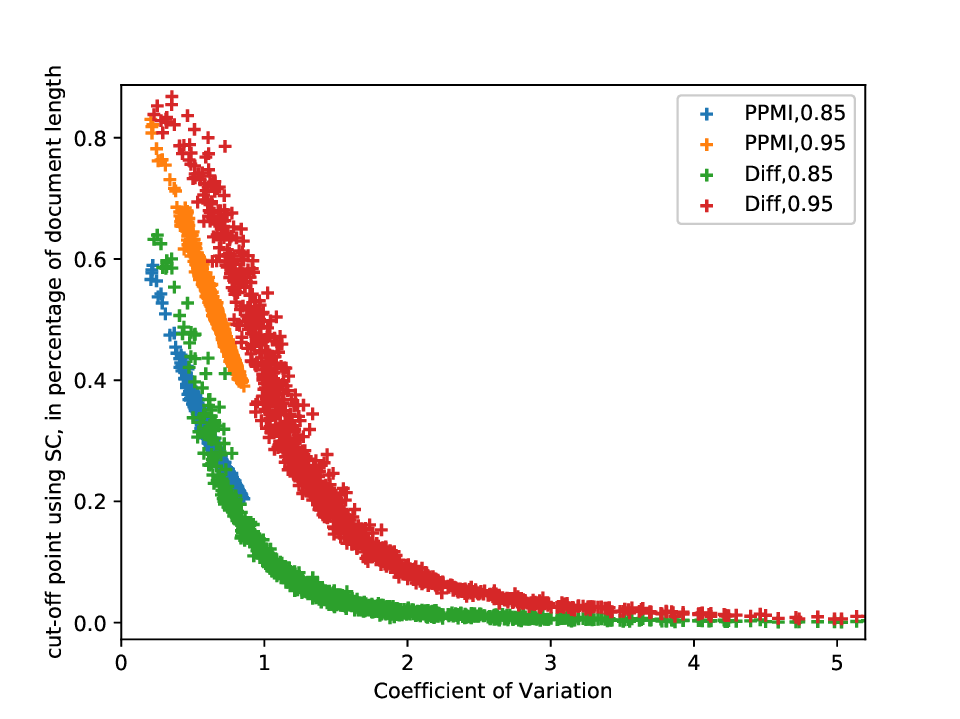}
}
\caption{  VT and SC functions  versus Coefficient of Variation}
%Izq CosenovsCoeficientVariacionPPMI.py
\label{figure:VTvsCV}
\end{figure}

We shall now attempt to analyze the relationship between concentration and cutoff point. In order to do so, we will focus on those cutoff functions that consider the distribution of the weights: VT\footnote{We do not consider the RC criterion since in our data the minimum weight value is almost zero and therefore the results for VT and RC are quite similar.} and SC. Figure \ref{figure:VTvsCV} illustrates the dependence of the cutoff-point in terms of the percentage of the document (y-axis) for the coefficient of variation (CV) on the x-axis. Focusing on  VT, top figures, we zoom  on low values of CV. More specifically, in the top left-hand corner we fixed the VT threshold in the figure to 50\% of the maximum weight in the ranking, and for each document, we plot its CV against the percentage of the document selected in the profile for the different weighting measures. In the top right-hand corner of the figure, we select PPMI and Diff as weighting measures and use as thresholds 25\% and 50\% of the maximum term weight. The bottom part of the figure shows the results obtained under similar circumstances using SC as the cutoff function, but in this case, we fix the SC threshold to 85\% and 95\% of the cosine similarity.

Various conclusions can be drawn from these figures. As expected, the cutoff point depends on the concentration of the distribution so that the greater the concentration, the smaller the profile. This seems to be true independently of the selected weighting measure and the selected cutoff function. Focusing on the results, TF, TFIDF and Diff seem to perform in a similar way since they exhibit similar trends, but PPMI performs differently, mainly because it generates fewer dispersed distributions. Consequently, more terms will be needed to represent the profile. For instance, when PPMI is used, 99.3\% of the profiles obtained using SC (with a threshold of 0.95) need more than 40\% of the regular terms, whereas this is true only for 23.1\% of the profiles obtained using Diff. The average sizes of the profiles when C-Col is used are 110.4, 145.4, 278.2 and 419.9 for Diff, TF, TFIDF and PPMI, respectively\footnote{Similar trends have been obtained for A-Col 104.7, 153.4, 400.0 and 818.1 and for I-Col 96.3, 105.1, 217.2 and 351.8 for Diff, TF, TFIDF and PPMI, respectively.}.

Finally, we would like to mention that each weighting measure will output a different ranking of terms and so the selected terms differ from each other. Whether or not this variability in the ranking is relevant for recommending purposes will be studied in Section \ref{section:system}. Nevertheless, at this point, we would like to highlight that Diff generally performs best and so we will focus on this weighting measure in the remainder of this section.

\begin{figure}[t]

%Izq ComparandoCosenoVT.py, Dcha ComparandoCosenoVT2.py,
\centering{
\includegraphics[width=0.60\textwidth]{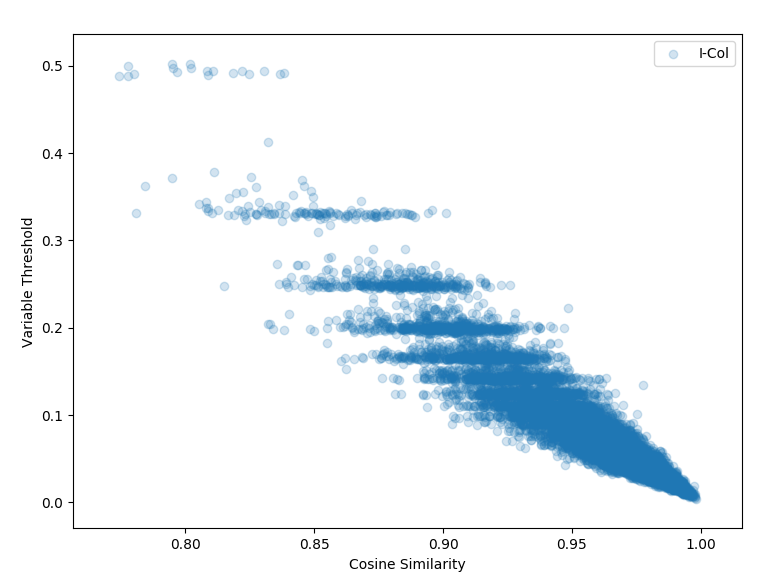}
}
\caption{Comparing Variable Threshold and Cosine Similarity criteria}
\label{figure:CSVT}
\end{figure}

We shall now focus on the comparison of VT and SC. Although there is generally a clear relationship with the concentration of the distribution, differences appear when we focus individually on each document, as Figure \ref{figure:CSVT} shows. For illustrative purposes, the scatterplot only focuses on the documents in I-Col, displaying the relationship between the SC (x-axis) and VT (y-axis) parameters for each document. More specifically, the values are obtained when a fixed percentage of terms (i.e. 50\%) is used in each document (the heaviest ones) and we plot the particular value of the parameter for both SC and VT that sets the cutoff to this particular point.  From this graph we can determine that there is a negative correlation between both metrics. We would, however, like to focus on the great variance between the SC scores that are related to a given VT value, as in the case of fixing VT at around 0.33. As expected, we can therefore confirm that VT is not able to capture properly the concentration 
of the weights in the distributions (which is also suggested by the differences between the graphs in Figure \ref{figure:VTvsCV}). This is due to the possible large value of the heaviest weight which could conceal a huge number of terms, even the most relevant ones. From now on, we will therefore focus on SC criterion.

\begin{figure}[t]

\centering{
\includegraphics[width=0.468\textwidth]{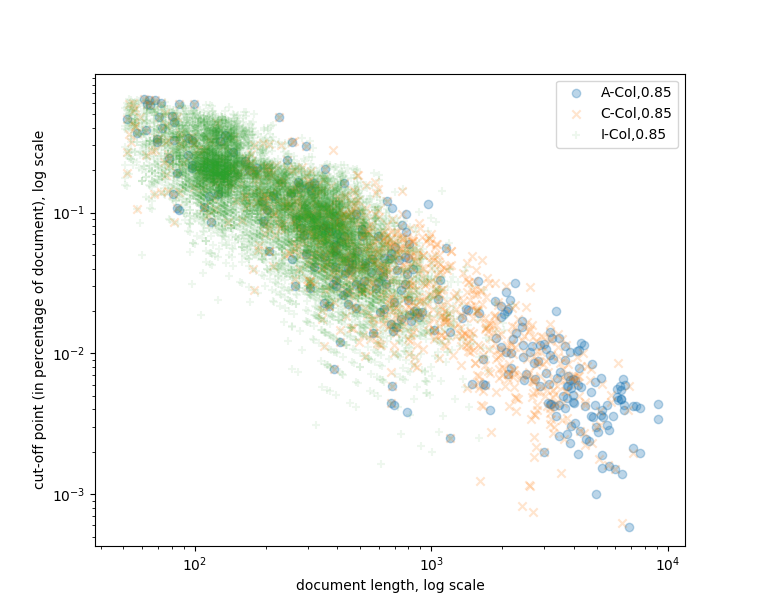}  \includegraphics[width=0.521\textwidth]{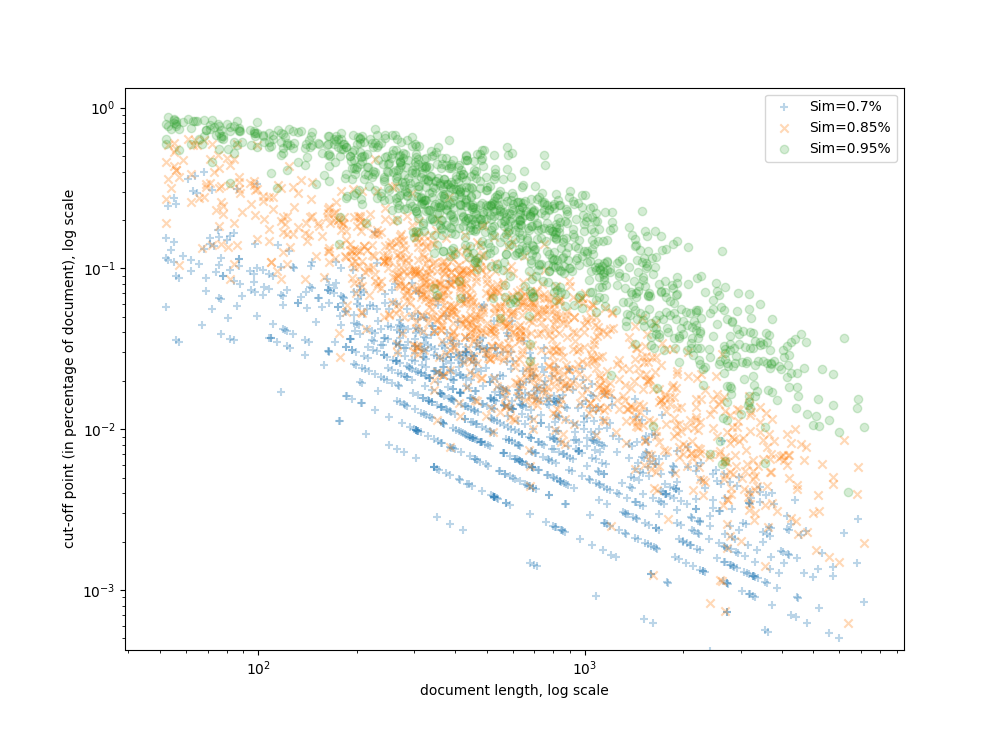}

}
\caption{Analyzing the performance of the SC function in terms of the document length for each s-collection (using a log-log scale)}
%Izq CosenovsTama.py, Dcha CosenovsTama2.py, \label{figure:CSvsSize}
\label{figure:CSvsSize_1}
\end{figure}

Continuing with the study of the size-related dependences, a different point of view is obtained if we fix the similarity threshold and analyze the value of the cutoff point (in terms of the percentage of the document). In this respect, if we look at Figure \ref{figure:CSvsSize_1}, it becomes apparent that the larger the document, the smaller the percentage of terms necessary to represent the same size distribution. The results also suggest a power law dependence. Going into the details, Figure \ref{figure:CSvsSize_1}     presents a   plot, using logarithmic scale on both axes to display not only the details of the small values but also the long-value trends. In these figures, each point is a particular document. In the left-hand side of Figure \ref{figure:CSvsSize_1} we fixed the similarity threshold to 0.85\% and present the log-log plot for every source collection, whereas in the right-hand side we focus on committee-based collection, C-Col\footnote{We choose C-Col because it is fairly representative as 
it has a relatively large number 
of documents of varying lengths.}, plotting the results for three different similarity thresholds of 0.70\%, 0.85\% and 0.95\%. In each case, the trend is for the points to lie approximately in a negatively sloping line (values around -1.0). This is mainly due to the fact that we obtain most of the distribution weight with a few terms and also that there is a large number of terms with small weights, with this long tail being greater for large documents. In order to illustrate this fact, Table \ref{table:raw} therefore displays the mean cutoff point value, i.e. the raw $l_j$, (and the standard deviation) when considering a similarity threshold of 0.70\%, 0.85\%, 0.95\%, 0.97\% and 0.99\%. We generally found that there are minimal differences (in terms of the cutoff point) among the source collections for lower thresholds. Only when we require a large amount of distribution (with thresholds above 0.95\%) is a lower number of terms (but also a higher percentage) needed for I-Col, whereas for C-Col and A-Col 
the cutoff remains the same.
%
% \begin{figure}[t]
%
%
% \centering{
% \includegraphics[width=0.45\textwidth]{Imagenes/CSvsSize-scol.png} \includegraphics[width=0.5\textwidth]{Imagenes/CSvsSize-ptj.png}
%
% }
% \caption{Analyzing the performance of the SC function in terms of the document length (using a log-log scale)}
% %Izq CosenovsTama.py, Dcha CosenovsTama2.py, \label{figure:CSvsSize}
% \label{figure:CSvsSize}
% \end{figure}

% 
% \begin{figure}[t]
% 
% 
% \centering{
%  \includegraphics[width=0.7\textwidth]{Imagenes/CSvsSize-ptj.png}
% 
% }
% \caption{Changing the similarity threshold for SC in C-Col (using a log-log scale)}
% %Izq CosenovsTama.py, Dcha CosenovsTama2.py, \label{figure:CSvsSize}
% \label{figure:CSvsSize}
% \end{figure}

\begin{table}[tbp]
	
	\caption{Mean cutoff point for the different collections }
	\label{table:raw}
\centering{\begin{tabular}{ cccc } \hline
 		& \textbf{A-Col} & \textbf{C-Col} & \textbf{I-Col} \\ \hline
\textbf{ 0.70\%} & $7.97 \pm 5.79$ & $8.13 \pm 4.99$ & $8.27 \pm 4.75$ \\
\textbf{ 0.85\%} & $26.85 \pm 16.68 $ & $ 27.85 \pm 15.53$ & $27.83 \pm 14.23$ \\
\textbf{ 0.95\%} & $104.75 \pm 54.74$ & $110.44 \pm 53.23$ & $ 96.28 \pm 41.96$ \\
\textbf{ 0.97\%} & $170.48 \pm 85.45$ & $174.25 \pm 82.34$ & $ 136.13 \pm 61.15$ \\
\textbf{ 0.99\%} & $361.37 \pm 198.32$ & $340.17 \pm 179.61$ & $207.45 \pm 111.01$ \\
\hline
\end{tabular}
}
\end{table}

\subsection{Analyzing unweighted-oriented cutoff functions}

\begin{figure}[tb]
%Izq ComparandoCosenoFN.py Dcha ComparandoCosenoVT.py
\centering{
\includegraphics[width=0.7\textwidth]{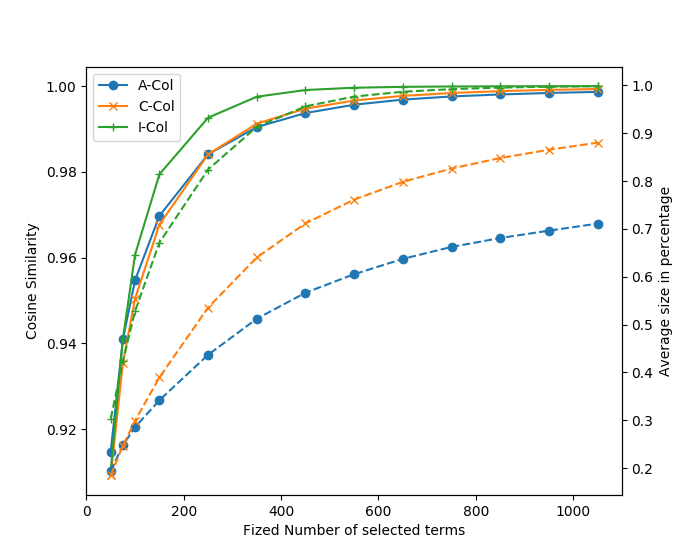}
\caption{Relating FN, FP and SC approaches}
\label{figure:CSvsVT_1}

}
\end{figure}

We will analyze the performance of the unweighted-oriented cutoff functions by  selecting a fixed number (FN) or considering  a percentage of terms in the document (FP) which help understand how the cutoff point varies according to the number of elements (terms) in the documents.  The average results for the different s-collections are summarized in Figure \ref{figure:CSvsVT_1}. The x-axis represents the case when the profiles were obtained when  FN varies from 50 to 1000. Two separate y-axes are used in this graph. The left y-axis shows the average value of the cosine similarity obtained for each FN (solid lines). From this plot, we can see that even for a small number of terms, a high cosine value was obtained (even for FN=50 we obtained cosine values over 91\%). Furthermore, the increase in cosine measure is not linear although it does depend on the number of terms in the documents. We can see that I-Col performs differently from C-Col and A-Col and this can be explained by the 
smaller number of terms in this type of document. This is demonstrated by the right y-axis (dashed lines) that represents the percentage of terms in the document (FP) for each FN, illustrating the variability in document length. From this data, we can see that although both A-Col and C-Col have a different number of terms, the most significant terms are located in higher positions in the ranking, and the addition of more terms to the profile, while   affecting the FP values, does not significantly affect SC.

\begin{figure}[tb]
%Izq ComparandoCosenoFN.py Dcha ComparandoCosenoVT.py
\centering{
 \includegraphics[width=0.7\textwidth]{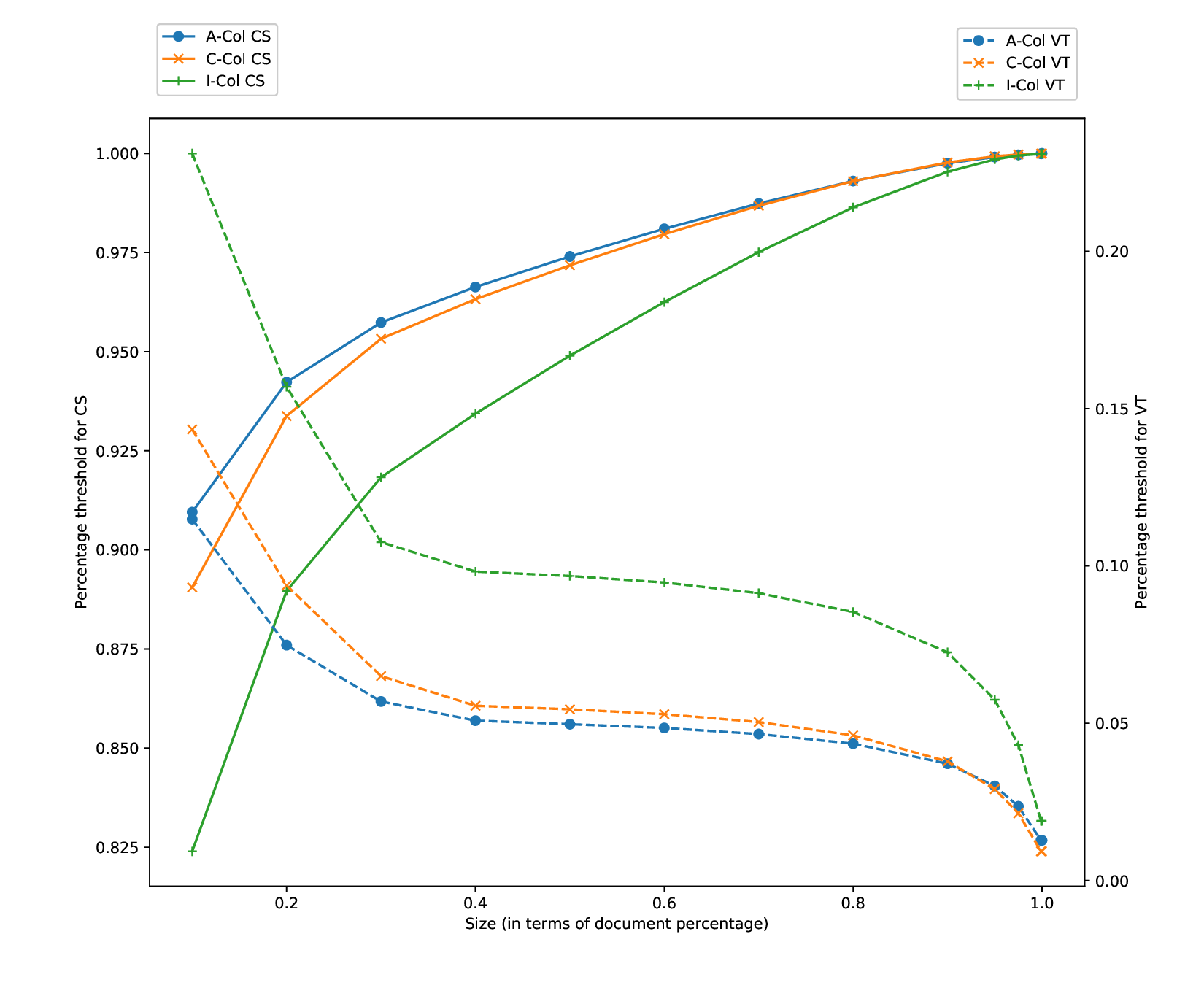}
\caption{Relating FP, VT and SC approaches.}
\label{figure:CSvsVT}

}

\end{figure}
We will now focus on the graph in Figure \ref{figure:CSvsVT} which represents how SC and VT depend on the FP. Once again, we show the mean values obtained for each s-collection. On the x-axis we represent the percentage of selected terms (rather than the full size), as Lorenz curves do. In terms of the y-axes, and taking into account the differences between the selected cutoff functions (SC and VT), we will also use two separate y-axes to represent the percentage of cosine measure obtained when $x_i\times 100 \%$ of items were considered for both $Sim(D^i_j,D_j)$ and $w_{t_i,j}/w_{t_1,j}$ and these are represented on the left- and right-hand side of the y-axis, respectively. It should be noted that for each FP value $x_i$, the y-axes represent the threshold values for SC and VT that set the cutoff to the point $l_j = x_i \times n_j$ (where $n_j$ is the size of the document $D_j$).

From these data we can conclude that both C-Col and A-Col behave in a similar manner. This might be due to the fact that in both cases each source document is obtained by aggregating different interventions into a single {\em virtual} document. This is not the case for I-Col, where each intervention is considered separately. Looking at the concentration curves for cosine measure (solid lines) we can therefore say that C-Col and A-Col are more concentrated than single interventions (I-Col). This might affect the cutoff point since if the similarity measure percentage is fixed, then the cutoff point ought to be greater for I-Col, or rather, a higher number of terms are needed to represent the same amount of document information. It is also interesting to note the skewness of the distributions in every case, and so with a relatively low number of terms (20\%) we can represent 90\% of the weights (for cosine measure). We can also observe that the number of terms with small weights is rather large and this is a 
further example of exponential distributions.

 In order to conclude this section, such reasoning suggests that when document-based profiles are considered, the cutoff point greatly depends on the size of the document and that with a relatively small number of terms we are able to capture the essence of the distribution. This situation could possibly justify the use of unweighted cutoff functions, such as for example by fixing the number (FN) or percentage (FP) of terms. Nevertheless, in Section \ref{sec:cv} we saw that the size of the document is not the only factor that affects the selected terms when using a weighted-based approach, since we obtained empirical findings to demonstrate that the concentration of the weight's distribution plays an important role. The existence of large concentrations of certain terms implies the rare occurrence of others and this can be considered as an indicator of document bias towards certain topics. The more unequal the distribution (more concentrated), the better we can exploit it to minimize profile size (by optimizing system efficiency). In the following section, we shall study the performance of the obtained profiles under the recommendation task, which is related to the quality of the selected terms.

\section{Performance  in the MP filtering task} \label{section:system}

Our system \cite{JIS16} builds MP profiles from the source collections. More specifically, after applying a cutoff function to determine the most representative terms, we transformed this set into a virtual document which will be indexed by an information retrieval system (IRS). To this end, it is necessary to convert each profile (i.e. a list of weighted terms) into a more standard representation of a document (a bag of words). In this paper, this transformation consists in   removing all the unselected terms in a source document\footnote{Another method which replicated terms proportionally to the computed weights was also proposed in \cite{JIS16}.}.

When a new document (e.g. a press release, a parliamentary initiative or a user's request) enters the system, it is used as a query and the IRS ranks the "documents" (i.e. the MP profiles) and recommends the new document to the top-ranked MPs. We will experiment with the cutoff functions and the weighting measures considered in Section \ref{sec:cutoff} and Section \ref{sec:w}, respectively. In order to measure the quality of the rankings, we use the well-known normalized discounted cumulative gain measure\cite{jarvelin02}, using only the first ten retrieved profiles (NDCG@10)\footnote{In \cite{JIS16} we also used other quality measures, e.g. Recall, Mean Average Precision and R-precision, but all these measures were shown to strongly correlate with NDCG@10.}.

%  and different methods to select the cutoff points: fixed number of terms (FN), fixed percentage (FP), variable threshold (VT), range-based term cutoff (RC) and the proposed similarity-based term cutoff (SC).

Various models can be used to match documents and profiles and these generally involve a weighted aggregation of the matching terms \cite{Alha17,baeza11,BFS12,HSC13}. Our approach was built using the Lucene library\footnote{http://lucene.apache.org}, and we selected its BM25 implementation since it is a state-of-the-art approach and performs better in terms of accuracy in preliminary experiments than other approaches such as, for example, the vector space model and language models, using their implementations in Lucene.

\subsection{Experimental setting} \label{section:experiments}

As in \cite{JIS16}, we used the three source collections, but since we want an accurate profile which truly represents the MP's interests, we have discarded any MP or technical guest who has participated in fewer than ten initiatives (thus keeping 132 MPs). The reduced collection will be named p-collection and in order to differentiate this from the original source collections, the learned profiles will be called monolithic profiles (M-Prof), committee-based subprofiles (C-SubP) and intervention-based subprofiles (I-SubP), and these are based on A-Col, C-Col and I-Col, respectively. Table \ref{table:ColSizes} shows the number of documents in each   p-collection, their average sizes in terms of unique keywords, and standard deviations.

\begin{table}[htbp]
	\centering
	\caption{Sizes of source collections when considering those MPs with more than 10 interventions}
	\begin{tabular}{rccc} \hline
			& \textbf{M-Prof} & \textbf{C-SupP} & \textbf{I-SubP} \\
\textbf{ Number of docs} & 132 & 949 & 9759 \\
\textbf{ Average size} & $3969.56 \pm 1720.9$ & $1078.87 \pm 1166.8$ & $281.34 \pm 219.6$ \\ \hline
	\end{tabular}
	\label{table:ColSizes}
\end{table}

We shall use a random subset of the initiatives for the training data, to build the different MP profiles. The remaining initiatives will be used for testing purposes, playing the role of documents to be recommended and introduced in the IRS as queries. We consider as ground truth the fact that a test initiative is only relevant to participating MPs. This conservative assumption\footnote{It is logical to suppose that an initiative might also be of interest to other MPs.} is necessary if we want to make use of available ``objective'' relevance judgments. We use the repeated holdout method for evaluation, i.e. this process is repeated five times and the results are averaged using 80\% and 20\% of the initiatives for training and testing, respectively.

The IRS will be fed with these p-collections of profiles. Given a query (i.e. a test initiative), the IRS will subsequently return a ranking of MP profiles that best match it. This MP ranking is direct if we are dealing with the monolithic collection as there is a one-to-one relationship between MPs and profiles, but this is not the case when working with subprofiles, as one MP could have many subprofiles. The original ranking obtained from a query for C-SubP consists of pairs (MP,Committee), and for I-SubP, of trios (MP,Initiative,Intervention). In order to obtain a final ranking of MPs, a fusion process must be performed. In this case, an MP's final score is computed by adding the different scores of their subprofiles while devaluating this value by considering their positions in the ranking, reduced logarithmically, as the discounted cumulative gain does in Information Retrieval evaluation \cite{jarvelin02}. By means of this devaluation, we penalize to a certain degree the occurrence of a subprofile in 
lower positions of the ranking.

\subsection{Analyzing weighting measures performance}

Our intention in this section is to study the quality of the terms selected by each weighting measure (Diff, TF, TFIDF and PPMI). With this purpose in mind, we decided to use FN as a cutoff function and fix the selected number of terms (FN) to the best 500 and 750 returned by each weighting metric so that all the metrics can be compared under the same conditions. The second and third columns in Table \ref{tab:ws} show the results obtained using the C-SupP collection, although similar results are obtained for the other collections. From these results, we can see that Diff achieves the best performance and that poor results are obtained for PPMI. This is because this metric is not able to capture the representativeness of a term from the point of view of a profile: infrequent terms which only appear in the document have a high PPMI value. Although these terms are specific to the document context, they do not therefore represent the user's interests as well as more frequent terms do.

\begin{table}[tbp]
\caption{Results for the different weighting measures when using  C-SubP.}
\begin{tabular}{l|r|r|r|r|r|r|r|r}
\hline
 & \multicolumn{1}{c|}{FN=500} & \multicolumn{1}{c|}{FN=750} & \multicolumn{3}{c|}{SC=95\%} & \multicolumn{3}{c }{SC=99\%} \\  
 & \multicolumn{1}{c|}{NDCG} & \multicolumn{1}{c|}{NDCG} & \multicolumn{1}{c }{NDCG} & \multicolumn{1}{c }{$l$} & \multicolumn{1}{c|}{ $\sigma$} & \multicolumn{1}{c }{NDCG} & \multicolumn{1}{c}{$l$} & \multicolumn{1}{c}{$\sigma$} \\ \hline
Diff & 0.701 & 0.702 & 0.589 & 106 & 55  & 0.670  & 325 & 187 \\  
TFIDF & 0.684  & 0.683 & 0.493 & 278 & 294 & 0.539  & 671 & 776  \\ 
TF & 0.680  & 0.676  & 0.587  & 145 & 67 & 0.622 & 447 & 265 \\ 
PPMI & 0.247 & 0.297  & 0.599  & 420 & 451 & 0.664 & 632 & 697 \\ \hline
\end{tabular}
\label{tab:ws}
\end{table}

\begin{figure}
 
\centering{
\includegraphics[width=0.7\textwidth]{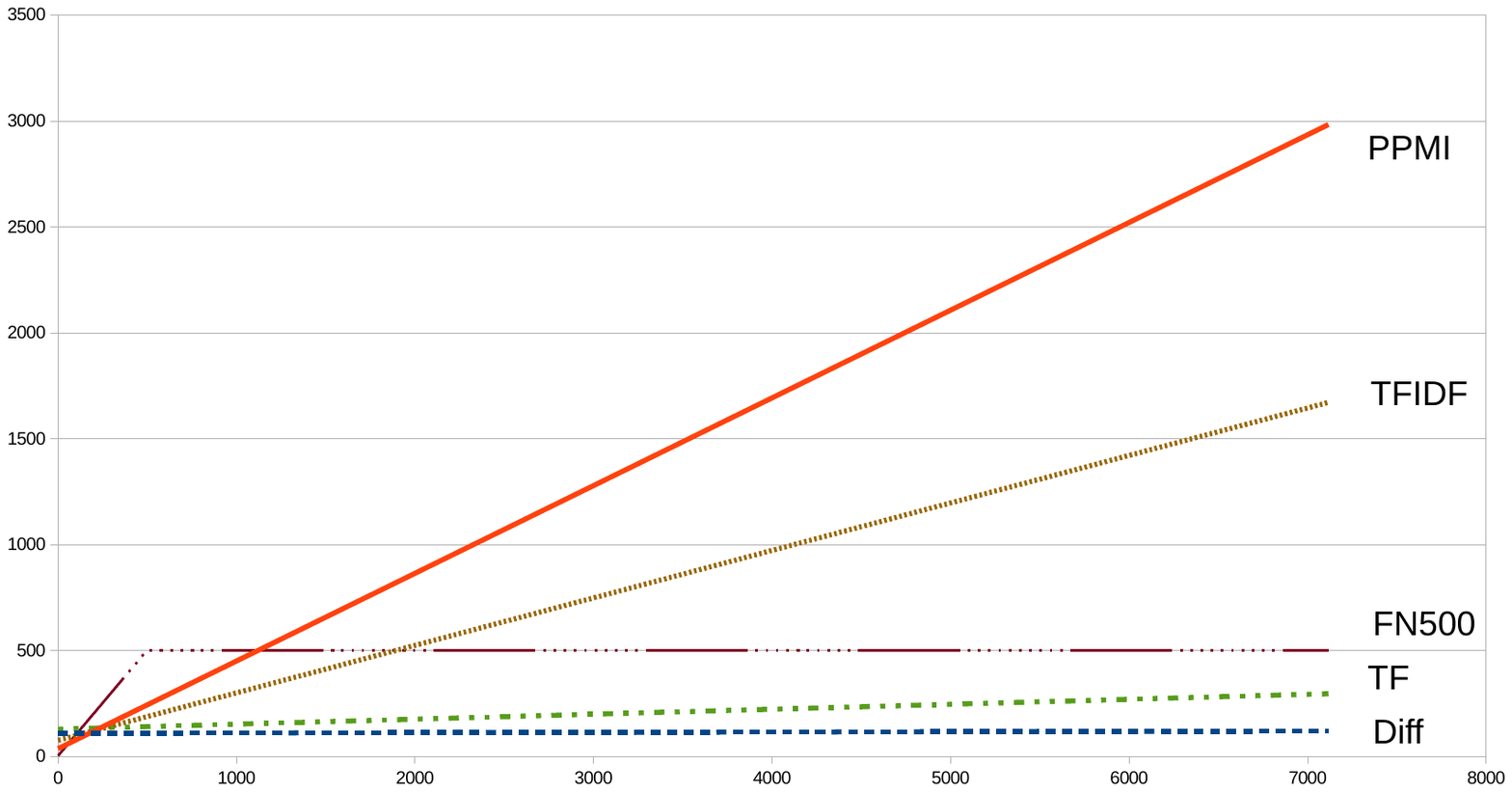}
}
\caption{Raw number of terms selected using SC for the different weighting metrics. }
\label{fig:raw_fn_vs_all}
\end{figure}

We also want to explore the performance of the proposed similarity-based term cutoff function (SC) when using different weighting metrics. We use two different thresholds to capture 95\% and 99\% of the similarity. For each one, Table \ref{tab:ws} shows the performance, NDCG@10, the average number of different terms in the profile, $l$, and its standard deviation, $\sigma$. The results show the good performance of Diff in terms of accuracy and number of selected terms: Diff seems to properly measure the importance of a term in terms of user interests and SC is also able to select the most relevant ones. The results might, however, be considered surprising since the combined use of the PPMI metric and SC selection criteria achieves fairly good results and so a deeper analysis is required. Table \ref{tab:ws} shows how these results have generally been obtained by using much larger profiles than those used with Diff (which needs smaller profiles with less variance). In order to better understand this performance, in Figure \ref{fig:raw_fn_vs_all} we show a scatterplot where the x-axis shows the size of the original profile (without selecting terms) and the y-axis shows the raw number of terms selected by SC (using a 95\% threshold) for each weighting metric. For comparison purposes, we also include the number of terms selected using FN500. In this case, it is apparent that both PPMI and TFIDF include a large number of terms when the original profile is large, which tallies with MPs with a large number of parliamentary speeches. For these MPs, we are therefore able to achieve good recommendations (particularly if we compare the results to those obtained using only the best 500 terms, FN = 500, for each weighting criterion) but at the expense of having much larger profiles.

\subsection{Analyzing cutoff functions performance}

From the previous analysis, Diff clearly performs well in terms of both the accuracy of the recommendations and efficiency (due to the small number of terms required) of the computations. In order to analyze the parameters associated with the different cutoff functions, in this section we will focus on the Diff weighting metric. We conducted preliminary experiments to obtain initial information about which parameter values produce reasonable results and we then searched for the best value in the selected ranges for each method. Table \ref{table:maximum} shows the accuracy of the different selection criteria under the recommendation task. The first row presents the results when no selection was made, i.e. we use the full profile, whereas for the remaining rows we present the optimum value for the parameter guiding the selection and accuracy obtained using the resulting subset of terms.

%We will attempt to understand the behavior of each individual selection method across collections.

\begin{table}[tbp]
	\centering
	\caption{Best parameter values for each selection method across p-collections and their NDCG@10 values}
	\begin{tabular}{rccc} \hline
 & \textbf{M-Prof}&\textbf{C-SubP}&\textbf{I-SubP} \\
 \hline
 \textbf{ full} & $0.7004 $ & $0.6849 $ & $0.6552$ \\ 		

\textbf{FN}&1725 / 0.7069 &750 / \bf{0.7024} &475 / 0.6669 \\
\textbf{FP}&38\% / 0.7126 &99\% / 0.6850 &99\% / 0.6526 \\
\textbf{RC}&0.25\%/ 0.7032 &0.525\% / 0.6955 &6.3\% / 0.6690 \\
\textbf{VT}&0.25\% / 0.7034 &0.5\% / 0.6946 &6\% / \bf{0.6724} \\
\textbf{SC}&99.925\% / \bf{0.7134} & 99.7\% / 0.6930 &95.9\% / 0.6606\\ \hline
	\end{tabular}
	\label{table:maximum}
\end{table}

Similar overall trends have been obtained for all of the criteria and p-collections, and these can be summarized as follows: when the number of selected terms is too small, accuracy results are not good. Nevertheless, as the number of terms in the profiles increases, the results improve until a maximum is reached. After this point, the addition of extra terms implies a reduction (albeit only slight) in performance. Consequently, any value of the parameters that ensures that a relatively high number of terms are selected enables us to obtain almost the best possible performance. This is because adding more terms than necessary has a minor impact on the final score of the documents. It should be noted that the retrieval models place more strength on the existence of matching terms between query and profile than on those without any match, i.e. terms in the profile but not in the query. 

If we focus on the comparison of selection methods, we could say that it is difficult to highlight the best one in absolute terms because the best performance in each p-collection is obtained by different techniques. We have computed t-tests among the selection techniques and chosen their best parameter values for each p-collection in an attempt to determine if we may discover whether one method is better than the others\footnote{The pattern that occurs in all the three p-collections is that there are no significant differences in the following pairs of selection methods: (FN, FP), (FP, RC) and (FP, VT). Additionally, for C-Prof and I-Prof, (SC, FP) do not show any significant differences, and for M-Prof and I-Prof, the same happens with (SC, RC). Positive significant differences are to be found in M-Prof from FN for RC and VT, from FN to VT, and SC from FP. In C-Prof, FN is statistically better than RC, VT and RC, and RC for VT. Finally, the patterns found in I-Prof show that there are positive significant 
differences from VT in RC and FN, RC with FN, and FN with SC.}. It is clear that there are no common patterns and significance   depends on the p-collection.

We would also like to point out that in order to achieve the best possible system performance for each collection, the values to assign to cutoff function parameters vary considerably and further experiments are needed to find these parameters. It is worth mentioning that there are two methods that perform well, although not the best, for the same parameter in every collection: FN with 750 terms and SC with a 99.7\% threshold, which results in an average accuracy loss of 1.76\% and 1.83\%, respectively, compared with the best results.

%FN using 750 terms (NDCG@10 values of 0.6890, 0.7024 and 0.6602, for M-Prof, C-SubP and I-SubP, respectively) and SC using  99.7\% of the cosine metric (NDCG@10 values of 0.6975, 0.6930 and 0.6553, for M-Prof, C-SubP and I-SubP, respectively).

% 
% Although accuracy is important, in this paper we wish to focus on the problem of how representative the profile is. More specifically, it is interesting to see that with smaller profiles we can obtain good recommendations (obtaining even better accuracy values than those obtained using the full profile) without a deterioration in efficiency (in terms of time) and obviously with smaller index sizes. This is an indication that the obtained profiles are a good representation of MP preferences.

Although accuracy is important, it might therefore be interesting to ascertain whether these methods, with their corresponding best values, are very demanding and build profiles close to the maximum possible size of terms, a fact that would clearly deteriorate performance in recommendation time. Due to obvious space restrictions, we have chosen one of the partitions\footnote{Although similar results are obtained for the other partitions.} and in Table \ref{table:occupancy} we show the occupancy percentages for each technique (100\% means that all the possible terms have been selected and included in the profile). We also show the number of profiles with an occupancy that is greater than or equal to 90\% of the total terms.

\begin{table}[tbp]
	\centering
	\caption{Percentage of occupancy of profiles for the p-collections and the best parameter values in partition A. In this table the rows represent: \%occ shows the average percentage of occupancy, $\sigma$(\%)occ represents the standard deviation, \#p the raw number of profiles with occupancy greater than or equal, and \%p the percentage of profiles with ocuppancy greater than or equal, respectively. }

\begin{tabular}{ rccccc } \hline
\multicolumn{6}{|c|}{{\bf M-Prof} – Avg. Prof. terms: $3969.56 \pm 1720.94$ }\\ \hline
& \bf{FN1725} & \bf{FP38} & \bf{RC0.250} & \bf{SC99.25} & \bf{VT0.250} \\
\%occ & 51.43 & 37.98 & 65.28 & 69.41 & 65.38\\
$\sigma$(\%)occ & 22.37 & 0.01 & 26.03 & 15.47 & 26.05\\
\#p $\geq$ 90\%& 12 & 0 & 31 & 4 & 32 \\
\%p $\geq$ 90\%& 9.09 & 0.0 & 23.48 & 3.03 & 24.24 \\
\hline
\multicolumn{6}{|c|}{{\bf C-SubP} – Avg. Prof. terms: $1078.87 \pm 1166.83$ }\\ \hline
 & {\bf FN750} & {\bf FP99} & {\bf RC0.525} & {\bf SC99.7} & {\bf VT0.5}\\
  \%occ& 77.76 & 98.69 & 87.39 & 72.84 & 88.59\\
$\sigma$(\%)occ & 30.16 & 2.13 & 21.61 & 21.59 & 20.96\\
\#p $\geq$ 90\% & 563 & 942 & 714 & 168 & 731\\
\%p $\geq$ 90\% & 59.39 & 99.37 & 75.32 & 17.72 & 77.11 \\
\hline
\multicolumn{6}{|c|}{{\bf I-SubP} – Avg. Prof. terms: $281.34 \pm 219.60$ }\\ \hline
 & {\bf FN475} & {\bf FP99} & {\bf RC6.3} & {\bf SC95.9} & {\bf VT6}\\
  \%occ & 96.35 & 98.11 & 59.81 & 51.23 & 68.23\\
$\sigma$(\%)occ & 11.10 & 4.23 & 33.75 & 21.94 & 34.46\\
\#p $\geq$ 90\% & 8681 & 9573 & 3295 & 288 & 4762\\
\%p $\geq$ 90\% & 88.95 & 98.09 & 33.76 & 2.95 & 48.80\\
\end{tabular}

	\label{table:occupancy}
\end{table}

The occupancy in M-Prof is halfway and there are very few profiles that contain at least 90\% of the selected terms. The FP method has the lowest percentage but it is important to remember that this is fixed for every profile, as occurs with FN. In C-SupP and I-SupP, the occupancy of the profiles increases considerably, as does the number of profiles with high occupancy. In these two p-collections, it is the SC method that keeps the smallest profiles and the fewest profiles with high occupancy with significantly better values than RC and VT. In M-Prof, although the percentage of profile occupancy is greater than the values presented by RC and VT (but in a similar order), the number of profiles with an occupancy greater than or equal to 90\% is much less. This is another interesting advantage of the SC method, which is able to keep the smallest profiles with equally high performance.

\section{Concluding remarks} \label{section:conclusions}

In this paper, we have presented a new approach to construct profiles based on the use of similarity metrics. We have conducted an axiomatic study of those properties that a selection function should fulfill. In order to perform this study we have drawn on those properties applied in a related field, i.e. the discrete concentration theory, to show that cosine measure satisfies most logical properties with the exception of the ``strongest'' property of the concentration theory, the transfer principle. Nevertheless, we show that a weaker variant of such a property is verified and that it is therefore a good selection approach. This theoretical study has been complemented with an empirical one, whereby we compare selection methods by considering real profiles in a parliamentary framework and have subsequently been used to recommend the most suitable politician when requested. %More specifically, the MPs profiles have been extracted from their speeches in committee or plenary sessions in the Andalusian 
Parliament and have subsequently been used to recommend the most suitable politician when requested.

We analyze the performance of five approaches that select a fixed number of terms (FN), a fixed percentage of terms (FP), a variable number of terms depending on a percentage of the term weights (RC and VT), and on a percentage of the admissible similarity with the full set of terms (SC), under three different test collections and using four different weighting metrics (TF, TFIDF, PPMI and Diff). 

We have examined the main two factors that affect the selection process: the number of terms and how the weights are distributed among them. From the experiments conducted and their results, we consider it much more appropriate to use a variable selection technique which adapts to the different sizes of the source documents.  We could conclude that Diff proves to be the best alternative for computing the weight of the terms, enabling good performance to be obtained in the MP search task with smaller profiles. With respect to the five selection methods, there is no clear winner, although it is true that SC is configured as a suitable approach for this selection task for a number of reasons. Firstly, it verifies most of the desired properties, naturally integrating both factors (size and weight distribution). Secondly, in terms of performance, it is very competitive. Furthermore, the range of percentages where this approach achieves the best results (while there are insignificant differences between other values) is really well bounded (around 99.7\%), irrespective of the s-collection used. Finally, since the profiles built with it are smaller, this in turn results in smaller indexes and faster recommendations.

In terms of recommendation, all the methods presented in this study rely on manually selecting the best value for the parameter (number of terms or percentage) by considering the s-collection from which the profiles will be built. This is problematic as it requires prior experimentation in order to determine the threshold for obtaining the optimal cutoff point with the subsequent waste of time and resources. In future lines of research, our aim will therefore be to design a method capable of computing the best value depending on the s-collection features.

Our next step  will be to apply clustering algorithms in order to automatically extract the MPs' topics of interest from their speeches and build the MP profiles with this information so that we are not limited by document discussion on a particular committee which might be biased by political criteria. Without doubt, eliminating this bias would be extremely beneficial.

\section*{ Acknowledgements}
This work has been funded by the Spanish Ministerio de Econom\'ia y Competitividad under projects TIN2013-42741-P and TIN2016-77902-C3-2-P, and the European Regional Development Fund (ERDF-FEDER).

\appendix

\section{Demonstrations}
\subsection{Cosine similarity satisfies the property of nominal increase: P5} \label{sec:demoP5}
{\em Principle of nominal increase:} If the weight of each term is increased by the same amount $h$, $h>0$, the distribution is less concentrated. By using econometric terminology the wealth is better distributed (the weights decrease in percentage) and so the cutoff point should not be decreased, i.e. $C(L_j) \leq C( L_j+h)$, where $ L_j+h =\{w_1+h, w_2+h, \ldots,w_n+h\}$.

\begin{pf}

We must prove that $C(L) \leq C(L + h)$, or equivalently that $ Sim((D+h)^i,(D+h))\leq Sim(D^i,D)$, for all $i \in \{ 1, \ldots, n\}$.

This is equivalent to proving that
\[ \frac{\sum_{k=1}^i (w_k+h)^2}{\sum_{k=1}^n(w_k+h)^2} \leq \frac{\sum_{k=1}^i w_k^2}{\sum_{k=1}^n w_k^2}\]

and after applying some basic manipulation this becomes
\[ (\sum_{k=1}^n w_k^2) (ih+ 2\sum_{k=1}^i w_k) \leq (nh + 2\sum_{k=1}^n w_k )(\sum_{k=1}^i w_k^2)\]
or equivalently
\[\frac{ih + 2\sum_{k=1}^i w_k}{\sum_{k=1}^i w_k^2} \leq \frac{nh+2\sum_{k=1}^n w_k}{\sum_{k=1}^n w_k^2}\]
%%%\[, \mbox{ i.e. }\frac{ih + 2\sum_{k=1}^i w_k}{\sum_{k=1}^i w_k^2} - \frac{hn +2\sum_{k=1}^n w_k}{\sum_{k=1}^n w_k^2}\leq 0\]
Since the right-hand side of the inequality does not depend on $i$, it can be considered constant. The equality holds trivially when $i=n$. It is therefore sufficient to demonstrate that $(ih + 2\sum_{k=1}^i w_k)/{\sum_{k=1}^i w_k^2}$ is an increasing function (in terms of $i$, i.e. the greater $i$, the greater its value). We can prove this by showing that
\[\frac{ih + 2\sum_{k=1}^i w_k}{\sum_{k=1}^i w_k^2} \leq \frac{(i+1)h + 2\sum_{k=1}^{i+1} w_k}{\sum_{k=1}^{i+1} w_k^2} \]

i.e.
\[(i h + 2\sum_{k=1}^i w_k)(w_{i+1}^2 + {\sum_{k=1}^{i} w_k^2} )\leq ((i+1)h + 2( w_{i+1} +\sum_{k=1}^{i} w_k) )(\sum_{k=1}^i w_k^2) \]
which after some manipulation becomes
\[ (i h + 2 \sum_{k=1}^i w_k) (w_{i+1}^2) \leq (h + 2 w_{i+1}) (\sum_{k=1}^i w_k^2) \]
and by grouping terms we obtain
\[ h (i w_{i+1}^2 - \sum_{k=1}^i w_k^2) \leq 2 w_{i+1} ( (\sum_{k=1}^i w_k^2) - (\sum_{k=1}^i w_k) w_{i+1}) \]

Since the weights are ordered decreasingly, $w_{i+1} \leq w_k, \forall k<=i$, and the weights are positive, we can see that $(i w_{i+1}^2 \leq \sum_{k=1}^i w_k^2)$ and also that $(\sum_{k=1}^i w_k^2) \geq (\sum_{k=1}^i w_k) w_{i+1}$, and therefore the inequalities are true. Consequently, P5 holds.

\end{pf}

\subsection{Cosine similarity satisfies the weak transfer principle: wP6}\label{sec:demoP6}

{\em Weak-transfer principle:} Given $ L_j =\{w_1,\ldots, w_n\}$ and two weights $w_a,w_b$ with $w_a > w_b$ and given $ L_j^+ =\{w_1,\ldots, w_a+h, \ldots,w_b-h, \ldots, w_n\}$ obtained by transferring an amount of weight from the lowest term, $w_b$, to the heaviest one, $w_a$. Let $l^+_a$ be the position of $w_a+h$ in $L_j^+$, then the weights are less concentrated for $l<l^+_a$ and more concentrated for $l\geq l^+_a$. Consequently, the cutoff point should increase before $l^+_a$ and decrease after $l^+_a$, i.e.

\begin{eqnarray}
\mbox{if } C(L_j)< l^+_a \mbox{ then } C(L_j) \leq C( L_j^+) \\
\mbox{if } C(L_j)\geq l^+_a \mbox{ then } C(L_j) \geq C( L_j^+).
\end{eqnarray}

\begin{pf} We will discuss the different situations that can be obtained when considering how $l^+_a$ relates to the cutoff point. In our experimentation, we rule out the subindex $j$. We will also use the subindex $m$ and the notation $\sum_{m=1}^{n\setminus \{a,b\}}$ to represent the fact that $m$ takes all the possible values in $\{1, \ldots, n\}$ with the exception of $a$ and $b$. Thus, for instance, $\sum_{m=1}^{n\setminus \{a,b\}} w_m = w_1 + w_2 + \ldots + w_{a-1} + w_{a+1} + \ldots + w_{b-1} + w_{b+1} + \ldots +w_n.$\\

$\bullet \mbox{ If } C(L )< l^+_a \mbox{ then } C(L ) \leq C( L^+):$ \\

In order to prove this property, we need to show that $Sim(D^{i+},D^+) \leq Sim(D^i,D)$, $ \forall i<l_a^+$, i.e.

\[ \frac{\sum_{k=1}^i w_{k}^2}{\sum_{m=1}^{n\setminus \{a,b\}} w_m^2 + (w_a + h)^2 + (w_b-h)^2} \leq \frac{\sum_{k=1}^i w_k^2}{\sum_{k=1}^n w_k^2}\]

We therefore only need to show that $\sum_{m=1}^{n\setminus \{a,b\}} w_m^2 + (w_a + h)^2 + (w_b-h)^2 \geq \sum_{k=1}^n w_k^2$, i.e. $\sum_{k=1}^n w_k^2 + (h^2 +2w_ah) +( h^2 -2w_b h) \geq \sum_{k=1} w_k^2$, which trivially holds since $w_a \geq w_b$. \\

\noindent$\bullet \mbox{ if } C(L)\geq l_a^+ \mbox{ then } C(L) \leq C( L^+):$\\

We must now see that $Sim(D^{i+},D^+) \geq Sim(D^i,D), \forall i \geq l_a^+$.

Firstly, let us assume that $q$ is an index value such that $l_a^+ \leq q < l_b^+$. We can therefore see that

\[ \frac{\sum_{m=1}^{q\setminus \{a\}} w_m^2 + (w_a + h)^2 }{\sum_{m=1}^{n\setminus \{a,b\}} w_m^2 + (w_a + h)^2 + (w_b-h)^2} \geq \frac{\sum_{k=1}^q w_k^2}{\sum_{k=1}^n w_k^2}\]

i.e.

\[ (\sum_{k=1}^q w_k^2 + (h^2 + 2w_ah) )(\sum_{k=1}^n w_k^2) \geq (\sum_{k=1}^n w_k^2 + (h^2 + 2w_ah) + (h^2 -2w_bh))(\sum_{k=1}^q w_k^2)\]

Since $(\sum_{k=1}^n w_k^2) \geq (\sum_{k=1}^q w_k^2)$, we only need to prove that $(h^2 + 2w_ah) \geq (h^2 + 2w_ah) + (h^2 -2w_bh)$, i.e. $2w_bh \geq h^2$. Taking into account that by definition $h \leq w_b$, this property holds.

Let us now assume that $r$ is an index value such that $r\geq l_b^+$. We will therefore see that

\[ \frac{\sum_{m=1}^{r \setminus \{a,b\}} w_m^2 + (w_a + h)^2 + (w_b-h)^2 }{\sum_{m=1}^{n\setminus \{a,b\}} w_m^2 + (w_a + h)^2 + (w_b-h)^2} \geq \frac{\sum_{k=1}^r w_k^2}{\sum_{k=1}^n w_k^2}\]

Once again, we must show that

 \[ (\sum_{k=1}^r w_k^2 + (h^2 + 2w_ah) + (h^2 -2w_bh) )(\sum_{k=1}^n w_k^2) \geq (\sum_{k=1}^n w_k^2 + (h^2 + 2w_ah) + (h^2 -2w_bh))(\sum_{k=1}^r w_k^2)\] which trivially holds since $(\sum_{k=1}^n w_k^2) \geq (\sum_{k=1}^r w_k^2)$.

\end{pf}
\subsection{Cosine similarity satisfies richest get richer inequality: P7}\label{sec:demoP7}
{\em Richest get richer inequality} states that if the weight of the highest-weighted term is increased by an amount $h$, $h>0$, the distribution is more concentrated, and therefore the cutoff point should not increase, i.e. $ C( L_j^*) \leq C(L_j)$, where $ L_j^* =\{w_1+h, w_2, \ldots,w_n\}$

\begin{pf}
We must prove that $C( L_j^*) \leq C(L_j)$, i.e. $ Sim(D^i,D) \leq Sim((D^*)^i,(D^*))$, for all $i \in \{ 1, \ldots, n\}$. We will therefore see that

\[ \frac{ (w_1 +h)^2 + \sum_{k=2}^{i} w_k^2 }{ (w_1 +h)^2 +\sum_{k=2}^{n} w_k^2 } \geq \frac{\sum_{k=1}^i w_k^2}{\sum_{k=1}^n w_k^2}\]

i.e.

\[ (h^2 + 2w_1 h + \sum_{k=1}^{i} w_k^2)( \sum_{k=1}^n w_k^2) \geq (\sum_{k=1}^i w_k^2)(h^2 + 2w_1 h + \sum_{k=1}^{n} w_k^2)\]
which trivially holds since $\sum_{k=1}^n w_k^2 \geq \sum_{k=1}^i w_k^2$.
\end{pf}

\end{document}